%
%
%
\magnification \magstep1
\hsize=16truecm
\vsize=22truecm
\voffset=0.4truecm
\hoffset=0.4truecm
\normalbaselineskip=5.25mm
\baselineskip=5.25mm
\parskip=5pt
\parindent=20pt
\nopagenumbers
\headline={\ifnum\pageno>1 {\hss\tenrm-\ \folio\ -\hss} \else {\hfill}\fi}
\def\em{\sl}
%
%
\newcount\EQNcount \EQNcount=1
\newcount\CLAIMcount \CLAIMcount=1
\newcount\SECTIONcount \SECTIONcount=0
\newcount\SUBSECTIONcount \SUBSECTIONcount=0
\def\actualnumber{\number\SECTIONcount}
\newcount\timecount
\def\TODAY{\number\day~\ifcase\month\or January \or February \or March \or
  April \or May \or June
  \or July \or August \or September \or October \or November \or December \fi
  \number\year\timecount=\number\time
  \divide\timecount by 60}
\newdimen\strutdepth
\def\DRAFT{\def\lmargin(##1){\strut\vadjust{\kern-\strutdepth
  \vtop to \strutdepth{
  \baselineskip\strutdepth\vss\rlap{\kern-1.2 truecm\eightpoint{##1}}}}}
  \font\footfont=cmti7
  \footline={{\footfont \hfil File:\jobname, \TODAY,  \number\timecount h}}}
\def\lmargin(#1){}
\def\ifundefined#1{\expandafter\ifx\csname#1\endcsname\relax}
\def\ifff(#1,#2,#3){\ifundefined{#1#2}
  \expandafter\xdef\csname #1#2\endcsname{#3}\else
  \write16{Warning : doubly defined #1,#2}\fi}
\def\NEWDEF #1,#2,#3 {\ifff({#1},{#2},{#3})}
\def\EQ(#1){\lmargin(#1)\eqno\tag(#1)}
\def\NR(#1){&\lmargin(#1)\tag(#1)\cr}  
\def\tag(#1){({\rm \actualnumber.\number\EQNcount})
  \NEWDEF e,#1,(\actualnumber.\number\EQNcount)
  \global\advance\EQNcount by 1
  }
\def\equ(#1){\ifundefined{e#1}$\spadesuit$#1\else\csname e#1\endcsname\fi}
\def\CLAIM #1(#2) #3\par{
  \vskip.1in\medbreak\noindent
  {\lmargin(#2)\bf #1~\actualnumber.\number\CLAIMcount.} {\sl #3}\par
  \NEWDEF c,#2,{#1~\actualnumber.\number\CLAIMcount}
  \global\advance\CLAIMcount by 1
  \ifdim\lastskip<\medskipamount
  \removelastskip\penalty55\medskip\fi}
\def\CLAIMNONR #1 #2\par{
  \vskip.1in\medbreak\noindent
  {\bf #1.} {\sl #2}\par
  \ifdim\lastskip<\medskipamount
  \removelastskip\penalty55\medskip\fi}
\def\clm(#1){\ifundefined{c#1}$\spadesuit$#1\else\csname c#1\endcsname\fi}
\def\sectionsize{\twelvepoint}
\def\sectiontype{\bf}
\newskip\beforesectionskipamount  
\newskip\sectionskipamount        
\beforesectionskipamount=24pt plus8pt minus8pt
\sectionskipamount=3pt plus1pt minus1pt
\def\sectionskip{\vskip\sectionskipamount}
\def\beforesectionskip{\vskip\beforesectionskipamount}
\def\SECTION#1\par{\vskip0pt plus.3\vsize\penalty-75
  \vskip0pt plus -.3\vsize
  \global\advance\SECTIONcount by 1
  \def\actualnumber{\number\SECTIONcount}
  \beforesectionskip\noindent
  {\sectionsize\sectiontype \actualnumber.\ #1}
  \EQNcount=1
  \CLAIMcount=1
  \SUBSECTIONcount=0
  \nobreak\sectionskip\noindent}
\def\SECTIONNONR#1\par{\vskip0pt plus.3\vsize\penalty-75
  \vskip0pt plus -.3\vsize
  \global\advance\SECTIONcount by 1
  \beforesectionskip\noindent
  {\sectionsize\sectiontype #1}
  \EQNcount=1
  \CLAIMcount=1
  \SUBSECTIONcount=0
  \nobreak\sectionskip\noindent}
\def\SECT(#1)#2\par{\lmargin(#1)
  \SECTION #2\par
  \NEWDEF s,#1,{\actualnumber}}
\def\sec(#1){\ifundefined{s#1}$\spadesuit$#1
  \else Section \csname s#1\endcsname\fi}
\def\subsectionsize{}
\def\subsectiontype{\bf}
\def\SUBSECTION#1\par{\vskip0pt plus.2\vsize\penalty-75
  \vskip0pt plus -.2\vsize
  \global\advance\SUBSECTIONcount by 1
  \beforesectionskip\noindent
  {\subsectionsize\subsectiontype \actualnumber.\number\SUBSECTIONcount.\ #1}
  \nobreak\sectionskip\noindent}
\def\SUBSECT(#1)#2\par{\lmargin(#1)
  \SUBSECTION #2\par
  \NEWDEF p,#1,{\actualnumber.\number\SUBSECTIONcount}}
\def\subsec(#1){\ifundefined{p#1}$\spadesuit$#1
  \else Subsection \csname p#1\endcsname\fi}
\def\APPENDIX(#1)#2\par{
  \def\actualnumber{#1}
  \SECTIONNONR Appendix #1. #2\par}
\def\REFERENCES{
  \parindent=30pt
  \parskip=5pt
  \SECTIONNONR References\par}

\def\PROOF{\medskip\noindent{\bf Proof.\ }}
\def\REMARK{\medskip\noindent{\bf Remark.\ }}
\def\REMARKS{\medskip\noindent{\bf Remarks.\ }}
\def\ACKNOWLEDGEMENTS{\medskip\noindent{\bf Acknowledgements.\ }}
\def\LIKEREMARK(#1){\medskip\noindent{\bf #1.\ }}
%
%
\let\endarg=\par
\def\finish{\def\endarg{\par\endgroup}}
\def\start{\endarg\begingroup}
\def\getNORMAL#1{{#1}}
\def\titlesize{\twelvepoint}
\def\titletype{\bf}
\def\TITLE{\beginTITLE\getTITLE}
  \def\beginTITLE{\start
     \titlesize\titletype\baselineskip=1.728
     \normalbaselineskip\rightskip=0pt plus1fil
     \noindent
     \def\endarg{\par\vskip.35in\endgroup}}
  \def\getTITLE{\getNORMAL}
\def\ENDTITLE{\endarg}
\def\AUTHOR{\beginAUTHOR\getAUTHOR}
  \def\beginAUTHOR{\start
    \vskip .25in\rm\noindent\finish}
  \def\getAUTHOR{\getNORMAL}
\def\FROM{\beginFROM\getFROM}
  \def\beginFROM{\start\parskip=0pt\vskip\baselineskip
    \def\finish{\def\endarg{\egroup\par\endgroup}}
    \vbox\bgroup\obeylines\eightpoint\em\finish}
  \def\getFROM{\getNORMAL}
\def\ABSTRACT#1\par{
  \vskip 1in {\noindent\sectionsize\sectiontype Abstract.} #1 \par}
\def\ENDABSTRACT{\vfill\break}
%
%
\newdimen\texpscorrection
\texpscorrection=0truecm  
\newcount\FIGUREcount \FIGUREcount=0
\newskip\ttglue
\newdimen\figcenter
\def\figure #1 #2 #3 #4\cr{\null
  \global\advance\FIGUREcount by 1
  \NEWDEF fig,#1,{Fig.~\number\FIGUREcount}
  \write16{ FIG \number\FIGUREcount: #1}
  {\goodbreak\figcenter=\hsize\relax
  \advance\figcenter by -#3truecm
  \divide\figcenter by 2
  \midinsert\vskip #2truecm\noindent\hskip\figcenter
  \includegraphics{#1}
  \vskip 0.8truecm\noindent \vbox{\eightpoint\noindent
  {\bf\fig(#1)}: #4}\endinsert}}
\def\figurewithtex #1 #2 #3 #4 #5\cr{\null
  \global\advance\FIGUREcount by 1
  \NEWDEF fig,#1,{Fig.~\number\FIGUREcount}
  \write16{ FIG \number\FIGUREcount: #1}
  {\goodbreak\figcenter=\hsize\relax
  \advance\figcenter by -#4truecm
  \divide\figcenter by 2
  \midinsert\vskip #3truecm\noindent\hskip\figcenter
  \includegraphics{#1}{\hskip\texpscorrection\input #2 }
  \vskip 0.8truecm\noindent \vbox{\eightpoint\noindent
  {\bf\fig(#1)}: #5}\endinsert}}
\def\fig(#1){\ifundefined{fig#1}$\spadesuit$#1
  \else\csname fig#1\endcsname\fi}
%
%
\catcode`@=11
\def\footnote#1{\let\@sf\empty 
  \ifhmode\edef\@sf{\spacefactor\the\spacefactor}\/\fi
  #1\@sf\vfootnote{#1}}
\def\vfootnote#1{\insert\footins\bgroup\eightpoint
  \interlinepenalty\interfootnotelinepenalty
  \splittopskip\ht\strutbox 
  \splitmaxdepth\dp\strutbox \floatingpenalty\@MM
  \leftskip\z@skip \rightskip\z@skip \spaceskip\z@skip \xspaceskip\z@skip
  \textindent{#1}\footstrut\futurelet\next\fo@t}
\def\fo@t{\ifcat\bgroup\noexpand\next \let\next\f@@t
  \else\let\next\f@t\fi \next}
\def\f@@t{\bgroup\aftergroup\@foot\let\next}
\def\f@t#1{#1\@foot}
\def\@foot{\strut\egroup}
\def\footstrut{\vbox to\splittopskip{}}
\skip\footins=\bigskipamount 
\count\footins=1000 
\dimen\footins=8in  
\catcode`@=12       
%
\font\twelverm=cmr12
\font\twelvei=cmmi12
\font\twelvesy=cmsy10 scaled\magstep1
\font\twelveex=cmex10 scaled\magstep1
\font\twelvebf=cmbx12
\font\twelvett=cmtt12
\font\twelvesl=cmsl12
\font\twelveit=cmti12
\font\ninerm=cmr9

\font\ninesy=cmsy9

\font\eightrm=cmr8
\font\eighti=cmmi8
\font\eightsy=cmsy8
\font\eightex=cmex8
\font\eightbf=cmbx8
\font\eighttt=cmtt8
\font\eightsl=cmsl8
\font\eightit=cmti8
\font\sixrm=cmr6
\font\sixi=cmmi6
\font\sixsy=cmsy6
\font\sixbf=cmbx6
\newfam\truecmr
\newfam\truecmsy
\font\twelvetruecmr=cmr10 scaled\magstep1
\font\twelvetruecmsy=cmsy10 scaled\magstep1
\font\tentruecmr=cmr10
\font\tentruecmsy=cmsy10
\font\eighttruecmr=cmr8
\font\eighttruecmsy=cmsy8
\font\seventruecmr=cmr7
\font\seventruecmsy=cmsy7
\font\sixtruecmr=cmr6
\font\sixtruecmsy=cmsy6
\font\fivetruecmr=cmr5
\font\fivetruecmsy=cmsy5
\textfont\truecmr=\tentruecmr
\scriptfont\truecmr=\seventruecmr
\scriptscriptfont\truecmr=\fivetruecmr
\textfont\truecmsy=\tentruecmsy
\scriptfont\truecmsy=\seventruecmsy
\scriptscriptfont\truecmsy=\fivetruecmsy
%
\def \eightpoint{\def\rm{\fam0\eightrm}
  \textfont0=\eightrm \scriptfont0=\sixrm \scriptscriptfont0=\fiverm
  \textfont1=\eighti  \scriptfont1=\sixi  \scriptscriptfont1=\fivei
  \textfont2=\eightsy \scriptfont2=\sixsy \scriptscriptfont2=\fivesy
  \textfont3=\eightex \scriptfont3=\eightex \scriptscriptfont3=\eightex
  \textfont\itfam=\eightit          \def\it{\fam\itfam\eightit}%
  \textfont\slfam=\eightsl          \def\sl{\fam\slfam\eightsl}%
  \textfont\ttfam=\eighttt          \def\tt{\fam\ttfam\eighttt}%
  \textfont\bffam=\eightbf          \scriptfont\bffam=\sixbf
  \scriptscriptfont\bffam=\fivebf   \def\bf{\fam\bffam\eightbf}%
  \textfont\truecmr=\eighttruecmr   \scriptfont\truecmr=\sixtruecmr
  \scriptscriptfont\truecmr=\fivetruecmr
  \textfont\truecmsy=\eighttruecmsy \scriptfont\truecmsy=\sixtruecmsy
  \scriptscriptfont\truecmsy=\fivetruecmsy
  \tt \ttglue=.5em plus.25em minus.15em
  \setbox\strutbox=\hbox{\vrule height7pt depth2pt width0pt}%
  \normalbaselineskip=9pt
  \let\sc=\sixrm  \let\big=\eightbig  \normalbaselines\rm
}
\def \twelvepoint{\def\rm{\fam0\twelverm}
\textfont0=\twelverm  \scriptfont0=\tenrm  \scriptscriptfont0=\eightrm
\textfont1=\twelvei   \scriptfont1=\teni   \scriptscriptfont1=\eighti
\textfont2=\twelvesy  \scriptfont2=\tensy  \scriptscriptfont2=\eightsy
\textfont3=\twelveex  \scriptfont3=\tenex  \scriptscriptfont3=\eightex
\textfont\itfam=\twelveit             \def\it{\fam\itfam\twelveit}%
\textfont\slfam=\twelvesl             \def\sl{\fam\slfam\twelvesl}%
\textfont\ttfam=\twelvett             \def\tt{\fam\ttfam\twelvett}%
\textfont\bffam=\twelvebf             \scriptfont\bffam=\tenbf
\scriptscriptfont\bffam=\eightbf      \def\bf{\fam\bffam\twelvebf}%
\textfont\truecmr=\twelvetruecmr      \scriptfont\truecmr=\tentruecmr
\scriptscriptfont\truecmr=\eighttruecmr
\textfont\truecmsy=\twelvetruecmsy    \scriptfont\truecmsy=\tentruecmsy
\scriptscriptfont\truecmsy=\eighttruecmsy
\tt \ttglue=.5em plus.25em minus.15em
\setbox\strutbox=\hbox{\vrule htwelve7pt depth2pt width0pt}%
\normalbaselineskip=12pt
\let\sc=\tenrm  \let\big=\twelvebig  \normalbaselines\rm
}
\catcode`@=11
\def\eightbig#1{{\hbox{$\textfont0=\ninerm\textfont2=\ninesy\left#1
  \vbox to6.5pt{}\right.\n@space$}}}
\catcode`@=12
%

\def\LL{{\cal L}}

\def\OO{{\cal O}}

\def\HB {\hfill\break}
\def\sqr#1#2{{\vcenter{\vbox{\hrule height .#2pt
        \hbox{\vrule width.#2pt height#1pt \kern#1pt
           \vrule width.#2pt}
        \hrule height.#2pt}}}}
\def\square{\mathchoice\sqr64\sqr64\sqr{2.1}3\sqr{1.5}3}
\def\QED{\hfill$\square$}

\def\real{{\bf R}}
\def\natural{{\bf N}}

\def\tsp{{\rm t}} 
\def\pp{{\rm a.e.}}
\def\phi{\varphi}
\def\ha{{\hat \alpha}}
\def\hb{{\hat \beta}}

\TITLE Stability of Travelling Waves for a Damped Hyperbolic Equation

\AUTHOR Th. Gallay and G. Raugel

\FROM CNRS et Universit\'e de Paris-Sud
      Analyse Num\'erique et EDP
      B\^atiment 425
      F-91405 Orsay, France
\ENDTITLE

\ABSTRACT
We consider a nonlinear damped hyperbolic equation in $\real^n$,
$1 \le n \le 4$, depending on a positive parameter $\epsilon$. If we set
$\epsilon=0$, this equation reduces to the well-known
Kolmogorov-Petrovski-Piskunov equation. We remark
that, after a change of variables, this hyperbolic equation has the same
family of one-dimensional travelling waves as the KPP equation. Using various
energy functionals, we show that, if $\epsilon >0$, these fronts are locally
stable under perturbations in appropriate weighted Sobolev spaces. Moreover,
the decay rate in time of the perturbed solutions towards the front of
minimal speed $c=2$ is shown to be polynomial. In the one-dimensional
case, if $\epsilon < 1/4$, we can apply a Maximum Principle for hyperbolic
equations and prove a global stability result. We also prove that the decay
rate of the perturbated solutions towards the fronts is polynomial, for all
$c > 2$.
\ENDABSTRACT

\SECTION Introduction

We consider the damped hyperbolic equation
$$
   \epsilon u_{tt}(\xi,t) + u_t(\xi,t) \,=\, \Delta_\xi u(\xi,t)
   + f(u(\xi,t))~, \EQ(u)
$$
where $\xi = (\xi_1,\xi_2,\dots,\xi_n) \in \real^n$, $t \in \real$,
$\epsilon > 0$, and $f : \real \to \real$ is a nonlinear map.
In the one-dimensional case $n=1$, equations of the form \equ(u)
arise as mathematical models describing various natural phenomena,
like the propagation of voltage along a nonlinear transmission line,
or the random motion of one-celled organisms [DO]. Here we consider
the multidimensional case also, and we are interested in situations
where the parabolic equation obtained by taking the limit $\epsilon \to 0$
in \equ(u) has a continuous family of travelling waves (or fronts)
propagating into the unstable state $u \equiv 0$. Sufficient conditions
on the nonlinearity $f$ for such a situation to occur are discussed for
example in Aronson and Weinberger [AW]. For convenience, we restrict
ourselves to the typical example of the Kolmogorov-Petrovsky-Piskunov equation
(KPP), which corresponds to $f(u) = u-u^2$, but more general nonlinearities
can be treated by the same methods.

Existence of travelling wave solutions to damped hyperbolic equations has
been proved by Hadeler [Ha] in a general context. In our case, this is
simply done by setting
$$
   u(\xi,t) \,=\, g(\sqrt{1+\epsilon c^2}\xi_1 - ct)~, \EQ(front)
$$
and inserting into \equ(u). One obtains for $g$ the differential equation
$$
   g''(x) + cg'(x) + g(x) - g(x)^2 \,=\, 0~. \EQ(g)
$$
It is well-known [AW] that, for all $c \ge 2$, this equation has a front-like
solution $g(x)$ satisfying $g'(x) < 0$ for all $x \in \real$, $\lim_{x \to
-\infty} g(x) = 1$, $\lim_{x \to +\infty} g(x) = 0$, and $g(x)$ is
unique up to a translation in the variable $x$. Therefore, for all
$\epsilon > 0$, Eq.\equ(u) has a continuous family of travelling waves
of the form \equ(front) indexed by the parameter $c \ge 2$. It should be
noted that the speed of such a wave is no longer $c$, but
$c/\sqrt{1+\epsilon c^2}$, a quantity which is bounded by $1/\sqrt{\epsilon}$
as $c \to \infty$. This is of course related to the finite propagation
speed property of equation \equ(u).

The stability of travelling waves for KPP and similar nonlinear parabolic
equations has been intensively studied over many years. Early results
have been obtained using comparison theorems based on the Maximum Principle,
see [KPP], [Fi], [AW]. Combined with probabilistic techniques, these
methods give a very detailed description of the basin of attraction of the
wave [Bn]. In parallel, a local stability analysis of the front
in suitable weighted spaces has been initiated by Sattinger [Sa] and
continued recently by Kirchg\"assner [Ki], Kapitula [Ka], Bricmont and
Kupiainen [BK], Gallay [Ga], Eckmann and Wayne [EW], using functional-analytic
techniques, renormalization group methods, or energy functionals.
In particular, the decay rate in time of the perturbations
in the critical case $c=2$ has been investigated [Ki], [BK], [Ga].
Similar results have also been obtained for higher dimensional equations
[MJ], and for systems of parabolic equations [KR], [RK].

The aim of this paper is to extend part of the stability results
available for the KPP equation to the hyperbolic equation \equ(u).
In particular, using energy functionals, we shall show that the
travelling waves \equ(front) are locally stable in appropriate function
spaces for all $c \ge 2$ and all $\epsilon > 0$. Moreover, using the
Maximum Principle for hyperbolic equations, we shall prove a global
stability result in the one-dimensional case, provided $\epsilon$
is sufficiently small. Finally, a decay rate as $t \to +\infty$ of the
perturbations will be obtained if $c=2$, or if $n=1$ and $\epsilon < 1/4$.

We now proceed to state our results in a more precise way. Given
$\epsilon > 0$, $c \ge 2$, we go to a moving frame using the change
of variables
$$
   u(\xi,t) \,=\, v(\sqrt{1+\epsilon c^2}\xi_1 - ct,\xi_2,\dots,\xi_n,t)
   \equiv v(x,y,t)~, \EQ(cv)
$$
where $x = \sqrt{1+\epsilon c^2}\xi_1 - ct$ and, if $n > 1$,
$y = (\xi_2,\dots,\xi_n)$. The equation for $v$ is
$$
   \epsilon v_{tt} + v_t - 2\epsilon c v_{xt} \,=\,
   v_{xx} + \Delta_y v + c v_x + v - v^2 ~, \EQ(v)
$$
and by construction $v(x,y,t) = g(x)$ is a stationary solution
of \equ(v). As in the parabolic case, this solution can only be stable
if we restrict ourselves to perturbations which decay to zero sufficiently
fast as $x \to +\infty$. To achieve this decay, we look for solutions of
the form $v(x,y,t) = g(x) + a(x)w(x,y,t)$, where $a(x) = e^{-\gamma x}$
for some $\gamma > 0$ which will be fixed later. Then $w(x,y,t)$
satisfies the equation
$$
   \epsilon w_{tt} + (1+2\epsilon c\gamma)w_t - 2\epsilon c w_{xt}
   \,=\, w_{xx} + \Delta_y w + (c-2\gamma) w_x + (1-c\gamma+\gamma^2-2g)w
   -aw^2~. \EQ(ww)
$$
Since Eq.\equ(ww) is of second order in time, we shall rewrite it
in the usual way as a first order system for the pair $(w,w_t)$, and
study the stability of the origin $(w,w_t)=(0,0)$ for this
system in a space $Z^1_\epsilon$ which we now describe.

\LIKEREMARK(Function spaces)For all $j \in \natural$, we denote by
$H^j = H^j(\real^n)$ the usual Sobolev space of order $j$ over
$\real^n$, with $H^0(\real^n) = L^2(\real^n)$. Similarly, we denote
by $H^j_a = H^j_a(\real^n)$ the weighted Sobolev space defined
by the norm $\|w\|_{H^j_a} = \|aw\|_{H^j}$. We also set $L^2_a = H^0_a$.
We write $X^j$ for the intersection $H^j \cap H^j_a$ equipped
with the norm $\|w\|_{X^j}^2 = \|w\|_{H^j}^2 + \|w\|_{H^j_a}^2$, and
$Z^j_\epsilon$ for the product $X^j \times X^{j-1}$ equipped with the
($\epsilon$-dependent) norm
$$
   \|(w_1,w_2)\|_{Z^j_\epsilon}^2 \,=\, \|w_1\|_{X^j}^2
   + \epsilon \|w_2\|_{X^{j-1}}^2~. \EQ(zjnorm)
$$
Finally, we define $Y_\epsilon = H^1 \times L^2$ and $Y_{\epsilon a} =
H^1_a \times L^2_a$, equipped with the ($\epsilon$-dependent) norms
$$
   \|(w_1,w_2)\|_{Y_\epsilon}^2 \,=\, \|w_1\|_{H^1}^2
   + \epsilon \|w_2\|_{L^2}^2~, \quad
   \|(w_1,w_2)\|_{Y_{\epsilon a}} \,=\, \|(aw_1,aw_2)\|_{Y_\epsilon}~.
$$
Remark that $Z^1_\epsilon \equiv Y_\epsilon \cap Y_{\epsilon a}$.

It follows from these definitions that $(w,w_t) \in Z^1_\epsilon$ if and
only if $(aw,aw_t)(1+e^{\gamma x}) \in H^1 \times L^2$. Therefore, our
perturbation space $\{(aw,aw_t) \,|\, (w,w_t) \in Z^1_\epsilon \}$ depends
on $\gamma$ and becomes smaller when $\gamma$ is increased. On the other
hand, using a direct calculation in Fourier space, it is not difficult to
verify that the origin in \equ(ww) is linearly stable in $Z^1_\epsilon$ if
and only if $1-c\gamma+\gamma^2 \le 0$. In fact, this condition can be
read off from the coefficient of $w$ in \equ(ww). As a consequence, the
biggest perturbation space in which we can hope for stability of the wave
is obtained by taking
$$
   \gamma \,=\, {c \over 2} - \sqrt{{c^2 \over 4} - 1}~. \EQ(gamma)
$$
Note that this value corresponds to the exponential decay rate of $g(x)$ as
$x \to +\infty$, since $g(x) \sim e^{-\gamma x}$ if $c > 2$ and
$g(x) \sim xe^{-x}$ if $c = 2$ [AW]. In the sequel, we shall always
assume that \equ(gamma) holds, so that \equ(ww) becomes
$$
   \epsilon w_{tt} + (1+2\epsilon c\gamma)w_t - 2\epsilon c w_{xt}
   \,=\, w_{xx} + \Delta_y w + \sqrt{c^2-4} \,w_x - 2gw -aw^2~.
   \EQ(w)
$$
Furthermore, we shall assume without loss of generality that $g(0) = 1-\sigma$
for some $\sigma \le 1/8$, and that $g(x) \ge 2a(x)/3$ for all $x \ge 0$.
This can always be achieved by replacing $g(x)$ by $g(x-x_0)$ for some
sufficiently large $x_0 > 0$.

\REMARK As in the parabolic case, one can show that the origin in \equ(ww)
is exponentially stable in $Z^1_\epsilon$ if $c > 2$ and $1 - c\gamma +
\gamma^2 < 0$. The fastest decay rate is obtained for the value
$$
   \hat \gamma(\epsilon) \,=\, {c \over 2} \sqrt{1+4\epsilon \over
   1+\epsilon c^2}~. \EQ(hatgam)
$$
Since these results are rather straightforward to prove, we shall focus
here on the marginal choice \equ(gamma) for which no exponential decay is
expected.

Using these definitions, we can state our first result, which shows that
the travelling waves are locally stable.

\CLAIM Theorem(local) Assume that $n \le 4$, and let $\epsilon_0 > 0$,
$c \ge 2$. Then there exist constants $\delta_0 > 0$ and $K_0 \ge 1$ such
that, for all $0 < \epsilon \le \epsilon_0$, the following holds~:
for all $(\phi_0,\phi_1) \in Z^1_\epsilon$ such that
$\|(\phi_0,\phi_1)\|_{Z^1_\epsilon} \le \delta_0$, there
exists a unique solution $(w,w_t) \in C^0(\real_+,Z^1_\epsilon)$ of
\equ(w) with initial data $(w(0),w_t(0))=(\phi_0,\phi_1)$. Moreover, one has
$$
   \|(w(t),w_t(t))\|_{Z^1_\epsilon} \,\le\, K_0
   \|(\phi_0,\phi_1)\|_{Z^1_\epsilon}~, \EQ(wwtbd)
$$
for all $t \ge 0$, and
$$
   \lim_{t \to +\infty} \left( \|\nabla w(t)\|_{X^0} + \|w_t(t)\|_{X^0}
   + \|w(t)\|_{L^2_a} \right) \,=\, 0~. \EQ(tozero)
$$
In addition, if $c=2$, one has
$$
   \lim_{t \to +\infty} \sqrt{t} \left( \|\nabla w(t)\|_{X^0}
   + \|w_t(t)\|_{X^0} + \|w(t)\|_{L^2_a} \right) \,=\, 0~. \EQ(decay)
$$

\REMARKS

\noindent 1.) By a solution of \equ(w), we always mean a {\sl mild} solution,
that is a solution of the integral equation associated with \equ(w),
see the proof of Proposition~2.1 below. In general, such solutions
satisfy \equ(w) in a distributional sense only, see Lions [Li], Section~1.1.
Remark that $w_{tt}$ or $w_{xx}$ belong to $C^0(\real_+,X^{-1})$ only, but
$w_{tt} - 2\epsilon c w_{xt} - w_{xx} - \Delta_y w \in C^0(\real_+,X^0)$
by \equ(w). Moreover, if $(\phi_0,\phi_1) \in Z^2_\epsilon$, then the
solution $(w,w_t)$ belongs to $C^1(\real_+,Z^1_\epsilon) \cap
C^0(\real_+,Z^2_\epsilon)$ and satisfies \equ(w) in a classical sense.
In \equ(wwtbd) and in the sequel, we use the short notation $w(t)$ for
$w(\cdot,\cdot,t)$, when no confusion is possible.

\noindent 2.) The restriction $n \le 4$ arises because we control
the nonlinearity $aw^2$ in \equ(w) in the energy space $Z^1_\epsilon$,
using the Sobolev embedding of $H^1(\real^n)$ into $L^4(\real^n)$.
More generally, if $f(u)$ in \equ(u) is a polynomial of degree $p > 1$,
we assume that $n \le 2p/(p-1)$. This bound could be improved up to
$2(p+1)/(p-1)$ using the more sophisticated $L_p-L_{p'}$ estimates of
Strichartz [St], [Br].

\noindent 3.) If $n \ge 3$, it follows from \equ(wwtbd), \equ(tozero)
and the Sobolev embedding theorem that
$$
   \lim_{t \to +\infty} \left(\|w(t)\|_{L^q} + \|aw(t)\|_{L^q}\right)
   \,=\, 0~, \quad 2 < q \le {2n \over n-2}~. \EQ(qdecay)
$$
If $n=2$, then \equ(qdecay) is valid for all $q > 2$ and even for
$q = \infty$ if $n=1$. In the case $c = 2$, \equ(wwtbd) and \equ(decay)
imply that
$$
   \lim_{t \to +\infty} \left(t^\eta \|w(t)\|_{L^q}
   + t^{1/2}\|aw(t)\|_{L^q}\right) \,=\, 0~, \quad \eta = {n(q-2) \over
   4q}~,
$$
for the same values of $q$.

\clm(local) is a local result in the sense that the size of the basin of
attraction of the wave, in particular its dependence on the parameter
$\epsilon > 0$, is not specified. However, in the parabolic limit
$\epsilon \to 0$, it is known [KR] that the travelling fronts are stable
with respect to large positive perturbations, and a similar phenomenon is
expected to hold for \equ(v) if $\epsilon$ is sufficiently small. To
investigate this, we restrict ourselves for convenience to one space
dimension, and we apply the Maximum Principle for hyperbolic equations,
which is briefly recalled in Appendix~A. Our second result reads:

\CLAIM Theorem(global) Assume that $n=1$, and let $\epsilon_0 > 0$,
$c \ge 2$, $d \in (0,1]$. Then for any $0 < \epsilon \le \epsilon_0$
and for any constant $K > 0$ such that
$$
   1 - 4 \epsilon (d+K) \,\ge\, 0~, \EQ(eps)
$$
there exists a constant $K^* = K^*(\epsilon_0,c,d,K) > 0$ such that the
following holds: for any $(\phi_0,\phi_1) \in Z^1_\epsilon$ satisfying
$\|(\phi_0,\phi_1)\|_{Z^1_\epsilon} \le K^*$ and, for (almost) every
$x \in \real$,
$$
   \phi_0(x) \,\ge\, - (1-d)a(x)^{-1}g(x)~, \EQ(phi0)
$$
$$
   \epsilon \phi_1(x) \,\ge\, \epsilon c \phi_0'(x) - {1 \over 2}
   (\phi_0 + (1-d)a^{-1}g)(x) + \epsilon c (-\gamma \phi_0 + (1-d)
   a^{-1}g')(x)~, \EQ(phi1)
$$
there exists a unique solution $(w,w_t) \in C^0(\real_+,Z^1_\epsilon)$
of \equ(w) with initial data $(\phi_0,\phi_1)$. Moreover, one has
$$
   \|(w(t),w_t(t))\|_{Z^1_\epsilon} \,\le\, K~, \quad
   w(x,t) \,\ge\, -(1-d)a^{-1}(x)g(x)~, \EQ(lwbd)
$$
for all $x \in \real$, $t \in \real_+$, and \equ(tozero), \equ(decay)
hold.

\REMARKS

\noindent 1) The proof will show that the constant $K^*$ can be chosen so
as to satisfy the equation $K_6 K^*(1+K^*)^{1/2} = K/2$ for some
$K_6 = K_6(\epsilon_0,c,d) \ge 1$. Therefore, if $\epsilon$ is small,
$K$ (hence $K^*$) can be chosen very big by \equ(eps), and
\clm(global) shows in this case that the travelling wave is stable
with respect to large perturbations, provided they satisfy the
``positivity conditions'' \equ(phi0), \equ(phi1).
Conversely, if $\epsilon$ is large, then $K$ (hence $K^*$) has to be
very small, and \clm(global) reduces to a local stability result similar
to \clm(local).

\noindent 2) The conditions \equ(phi0), \equ(phi1) appear when applying
the Maximum Principle to the equation \equ(v), see Appendix~A.
The first one simply says that $v(x,0) \ge d g(x)$ for all $x \in \real$.
The condition on the derivative is not very restrictive if $\epsilon$ is
small, and disappears in the limit $\epsilon \to 0$.

The previous results are incomplete in the sense that they fail to
give a decay rate for the perturbations when $c > 2$. Also, it would
be very natural to have at least a global existence result if $d = 0$.
Indeed, it is known that, if $0 \le v(x,0) \le g(x)$, the solution $v(x,t)$
of the parabolic equation \equ(v) with $\epsilon = 0$ exists for all times
and satisfies $0 \le v(x,t) \le g(x)$. In terms of the variable $w$, this
corresponds to $-a^{-1}(x)g(x) \le w(x,t) \le 0$. A similar
property is expected to hold for the hyperbolic equation \equ(w) if
$\epsilon$ is sufficiently small.

A partial answer to these two questions can be given when $\epsilon \le 1/4$.
Indeed, in this case the Maximum Principle allows us to compare the solution
$w(x,t)$ of \equ(w) with solutions of {\sl linear} equations, whose initial
data are the ``positive'' and ``negative'' parts $(\phi_0^\pm,\phi_1^\pm)$
of $(\phi_0,\phi_1)$, in the sense of Appendix~A. They are given by
$$ \eqalign{
   \phi_0^+(x) &= \sup(0,\phi_0(x))\,, \cr
   \phi_1^+(x) &= c (\phi_0^+)'(x) - ({1 \over 2\epsilon}+c\gamma)
     \phi_0^+(x) + \sup(0,(\phi_1-c\phi_0' + ({1 \over 2\epsilon}+c\gamma)
     \phi_0)(x))\,,} \EQ(phi+)
$$
and
$$ \eqalign{
   \phi_0^-(x) &= \inf(0,\phi_0(x))\,, \cr
   \phi_1^-(x) &= c (\phi_0^-)'(x) - ({1 \over 2\epsilon}+c\gamma)
     \phi_0^-(x) + \inf(0,(\phi_1-c\phi_0' + ({1 \over 2\epsilon}+c\gamma)
     \phi_0)(x))\,.} \EQ(phi-)
$$
Remark that $(\phi_0^\pm,\phi_1^\pm)$ belong to $Z^1_\epsilon$ and that
$\phi_i = \phi_i^+ + \phi_i^-$ for $i = 0,1$. Although the norms of
$\phi_1^\pm$ seem to depend strongly on $\epsilon$, it is not the case
actually: the reader may check that $|\phi_1^\pm(x)| \le |\phi_1(x)| +
c |\phi_0'(x)|$ {\pp} in $\real$.

With these definitions, we can state the last result:

\CLAIM Theorem(decay) Assume that $n=1$, and let $c \ge 2$, $d \in [0,1]$.
Then for any $0 < \epsilon \le 1/4$ and for any nonnegative constant $K$
satisfying
$$
   1 - 4 \epsilon (1+K) \,\ge\, 0~,
$$
there exists a nonnegative constant $\tilde K = \tilde K(c,K)$ such that the
following holds: for any $(\phi_0,\phi_1) \in Z^1_\epsilon$ satisfying
\equ(phi0), \equ(phi1) and
$$
   \inf\left(\|(\phi_0,\phi_1)\|_{Z^1_\epsilon}\,,\,
   \|(\phi_0^+,\phi_1^+)\|_{Z^1_\epsilon}\right) \,\le\, \tilde K~,
$$
where $(\phi_0^+,\phi_1^+)$ is given by \equ(phi+), there exists a unique
solution $(w,w_t) \in C^0(\real_+,Z^1_\epsilon)$ of \equ(w) with initial
data $(\phi_0,\phi_1)$. Moreover, one has
$$
   -(1-d)g(x) \,\le\, a(x)w(x,t) \,\le\, K~,
$$
for all $x \in \real$, $t \in \real_+$. Finally, if $d > 0$ and
$\epsilon < 1/4$, one has
$$
   \lim_{t \to +\infty} t^{1/4} \left(\|w(t)\|_{L^\infty} +
   \|(w(t),w_t(t))\|_{Y_{\epsilon a}} \right) \,=\, 0~.
$$

\REMARK The constant $\tilde K$ is given by $\tilde K = K/N$,
where $N = N(c)$ is a positive constant. Note that the case $K = 0$
is non trivial: it corresponds to nonpositive initial data.

An outline of the paper is as follows. In Section~2 we introduce energy
functionals which allow us to derive {\sl a priori} estimates for the
solutions $w(x,t)$ of \equ(w) under the assumption that either $\|w(t)\|_{X^0}$
is sufficiently small or $w(x,t)$ satisfies the lower bound in \equ(lwbd)
on some time interval. Using these energy estimates, we prove \clm(local)
in Section~3. Section~4 is devoted to the one-dimensional case $n=1$.
Combining the Maximum Principle with the estimates of Section~2, we
derive \clm(global). Furthermore, when $\epsilon \le 1/4$,
we obtain linear bounds for the solutions of \equ(w) which allow us
to prove \clm(decay). In Section~5, we consider the limiting parabolic
equation \equ(u) when $\epsilon = 0$. Noting that all the estimates
made in Section~2 are uniform in $\epsilon$ when $0 < \epsilon \le
\epsilon_0$, and using the Maximum Principle for parabolic equations,
we obtain analogues of \clm(local) and \clm(decay). Thereby we recover
some known stability results for the KPP equation.
Finally, in Appendix~A, we recall the Maximum Principle for hyperbolic
equations [PW] in a version adapted to our purposes.

\SECTION Energy Estimates

In this section, we derive some {\sl a priori} estimates for the solutions
of \equ(w) which will be needed in the proofs of \clm(local) and
\clm(global). We begin with a standard local existence result.

\CLAIM Proposition(loc-ex) Let $\epsilon > 0$, $c \ge 2$, and let
$(\phi_0,\phi_1) \in Z^1_\epsilon$. Then there exists a time $T =
T(\epsilon,c,\phi_0,\phi_1) > 0$ such that \equ(w) has a unique
solution $(w,w_t) \in C^0([0,T],Z^1_\epsilon)$ satisfying $(w(0),w_t(0))
= (\phi_0,\phi_1)$.

\REMARK In fact, the proof gives a lower bound on the existence time
which depends only on $\epsilon,c$ and $\|(\phi_0,\phi_1)\|_{Z^1_\epsilon}$.
Moreover, the energy estimates below will show that this time is
independent of $\epsilon$ if $\epsilon \in (0,\epsilon_0]$.

\PROOF Setting $W = (w,w_t)^\tsp$ (where ${}^\tsp$ denotes the transposition),
we rewrite \equ(w) into the ``abstract form''
$$
   \dot W \,=\, AW + F(W)~, \EQ(abs)
$$
where $A$ is the linear operator
$$
   A \,=\, \pmatrix{ 0 & 1 \cr
   \epsilon^{-1}(\partial_x^2 + \Delta_y + \sqrt{c^2-4}\partial_x -2g)
   & -\epsilon^{-1} + 2c(\partial_x-\gamma) \cr}~,
$$
and $F(W) = (0,-\epsilon^{-1}aw^2)^\tsp$. It is not difficult to show
that the operator $A$, defined on the domain $D(A) = Z^2_\epsilon$,
is the generator of a $C^0$-semigroup [Pa] of bounded linear operators
in $Z^1_\epsilon$. Indeed, $A$ can be written as the sum of a bounded
operator (depending on $x \in \real$ through the function $g$) and an
unbounded operator with constant coefficients, for which the property
of being a generator can be verified by a direct
calculation (using Fourier transforms). Therefore, it follows from a
classical stability theorem ([Pa], Theorem~3.1.1) that $A$ is the
generator of a $C^0$-semigroup $e^{At}$ in $Z^1_\epsilon$.

On the other hand, it is easy to verify that $F : Z^1_\epsilon \to
Z^1_\epsilon$ is a $C^1$ map. Indeed, if $w \in X^1 = H^1 \cap H^1_a$,
then by the Sobolev embedding theorem $\|w\|_{L^4}^2 \le K_S \|w\|_{H^1}^2$
and $\|aw\|_{L^4}^2 \le K_S \|w\|_{H^1_a}^2$ for some $K_S > 0$. Therefore,
$\|aw^2\|_{L^2}^2 \le \|aw\|_{L^4}^2 \|w\|_{L^4}^2 \le K_S^2 \|w\|_{H^1}^2
\|w\|_{H^1_a}^2$ and $\|aw^2\|_{L^2_a}^2 = \|aw\|_{L^4}^4 \le
K_S^2 \|w\|_{H^1_a}^4$. Combining these inequalities, we find that
$\|aw^2\|_{X^0} \le K_S \|w\|_{X^1}^2$, which proves that
$F$ maps $Z^1_\epsilon$ into itself. Since $F$ is quadratic, the
differentiability follows by the same estimates.

In view of these properties, a standard result in semigroup theory ([Pa],
Theorem~6.1.4) implies that, for all $\Phi = (\phi_0,\phi_1)
\in Z^1_\epsilon$, the integral equation
$$
   W(t) \,=\, e^{At} \Phi + \int_0^t e^{A(t-\tau)} F(W(\tau))\,d\tau~,
$$
has a unique solution $W \in C^0([0,T],Z^1_\epsilon)$ for some $T > 0$.
This is what we call a (mild) solution of \equ(abs), hence of \equ(w).
Moreover, if $\Phi \in Z^2_\epsilon$, then $W \in C^1([0,T],Z^1_\epsilon)
\cap C^0([0,T],Z^2_\epsilon)$ and satisfies \equ(abs) in a classical
sense ([Pa], Theorem~6.1.5). \QED

In the sequel, we fix $\epsilon_0 > 0$, $c \ge 2$, and for some $\epsilon \in
(0,\epsilon_0]$ we assume that we are given a solution $W = (w,w_t)$ of
\equ(w) (in the sense of \clm(loc-ex)) defined on some time interval $[0,T]$
and satisfying {\sl one} of the following two assumptions:

\noindent{\bf Hypothesis H1:} \HB
\centerline{$\sup \{\|w(t)\|_{X^0} \,|\, t \in [0,T]\} \le
\delta$ for some sufficiently small $\delta > 0$,}

\noindent{\bf Hypothesis H2:} \HB
\centerline{$w(x,y,t) \ge -(1-d)a(x)^{-1}g(x)$ $\pp(x,y)$,
$\forall t \in [0,T]$, for some $d \in (0,1]$.}

These two cases are adapted to the purposes of the proofs of \clm(local)
and \clm(global) respectively. To be specific, we assume in the first case
that $\delta \le 1/(8K_S)$, where $K_S$ is the constant of the Sobolev
embedding of $H^1$ into $L^4$ (like in the proof of \clm(loc-ex)).

Under these assumptions, we shall study two families of energy functionals:
{\sl unweighted} and {\sl weighted} ones, which control the size of the
solution $w(x,y,t)$ in the spaces $Y_\epsilon$ and $Y_{\epsilon a}$
respectively. We shall derive differential inequalities for these
functionals, which will show that the solution $w(x,y,t)$ is bounded
uniformly in time by a quantity depending only on the initial data.

\SUBSECTION Unweighted Functionals

Given $w(x,y,t)$ as above, we define
$$ \eqalign{
   E_0(t) \,&=\, \int_{\real^n} \left({\epsilon \over 2}w_t^2 +
     {1 \over 2}|\nabla_z w|^2 + gw^2 + {1 \over 3}aw^3 \right) dz~, \cr
   E_1(t) \,&=\, \int_{\real^n} \left( {1+2\epsilon c\gamma \over 2} w^2 +
     \epsilon ww_t \right)\,dz~, \cr
   E_2(t) \,&=\, \alpha E_0(t) + E_1(t)~,\phantom{1 \over 2}} \EQ(e0)
$$
where $\alpha = \max(2\epsilon,1/(2c^2))$. Here and in the sequel, we
set $z = (x,y) \in \real^n$ and $dz = dxdy$.

\CLAIM Lemma(e0) Assume that H1 or H2 holds. Then
$$
   E_0(t) \,\ge\, \int_{\real^n} \left({\epsilon \over 2}w_t^2
   + {1 \over 4}|\nabla_z w|^2 + {1 \over 2}gw^2 \right)dz~, \EQ(bde0)
$$
for all $t \in [0,T]$. Moreover, $E_0 \in C^1([0,T])$ and
$$
   \dot E_0(t) \,\le\, (c^2-4) E_0(t)~. \EQ(dote0)
$$

\PROOF We first control the cubic term in \equ(e0). Using the Cauchy-Schwarz
and Sobolev inequalities, we have
$$ \eqalign{
   \left| \int_{x\le 0} aw^3\,dz \right| \,&\le\,
     \left(\int_{x\le 0} a^2 w^2\,dz\right)^{1/2}
     \left(\int_{x\le 0} w^4\,dz\right)^{1/2} \cr
   \,&\le\, K_S \|w\|_{X^0} \int_{x\le 0} \left(w^2 + |\nabla_z w|^2
     \right)\,dz \cr
   \,&\le\, 2 K_S \|w\|_{X^0} \int_{x \le 0} \left(gw^2 + {1 \over 2}
     |\nabla_z w|^2 \right)\,dz~,} \EQ(xneg)
$$
since $g(x) \ge 1/2$ for $x \le 0$. Similarly,
$$ \eqalign{
   \left| \int_{x\ge 0} aw^3\,dz \right| \,&\le\,
     \left(\int_{x\ge 0} w^2\,dz\right)^{1/2}
     \left(\int_{x\ge 0} a^2 w^4\,dz\right)^{1/2} \cr
   \,&\le\, K_S \|w\|_{X^0} \int_{x \ge 0} \left(aw^2 + |\nabla_z(a^{1/2}w)|^2
     \right)\,dz~.}
$$
The integral of $|\nabla_z(a^{1/2}w)|^{1/2}$ is equal to
$$
   \int_{x\ge 0}\left(a|\nabla_z w|^2 + {\gamma^2 \over 4}aw^2 - \gamma
   aww_x \right)\,dz
   \,\le\, \int_{x\ge 0} \left({3 \over 4}\gamma^2 a w^2 + {3 \over 2}
     a|\nabla_z w|^2 \right)\,dz~.
$$
Since $\gamma^2 \le 1$ and $2a(x)/3 \le g(x) \le 1$ for $x \ge 0$, we thus
have
$$ \eqalign{
   \left| \int_{x\ge 0} aw^3\,dz \right| \,&\le\,
     K_S \|w\|_{X^0} \int_{x \ge 0} \left(2aw^2 + {3 \over 2} a|\nabla_z w|^2
     \right) \,dz\cr
   \,&\le\, 3 K_S \|w\|_{X^0} \int_{x\ge 0} \left(gw^2 + {1 \over 2}
     |\nabla_z w|^2 \right)\,dz~.} \EQ(xpos)
$$
Combining \equ(xneg) and \equ(xpos), we conclude
$$
   \left| \int_{\real^n} aw^3 \,dz \right| \,\le\, 3 K_S \|w\|_{X^0}
   \int_{\real^n} \left(gw^2 + {1 \over 2} |\nabla_z w|^2 \right)\,dz~.
   \EQ(aw3)
$$
If H1 holds, one has $3K_S \|w\|_{X^0} \le 3K_S \delta \le 1/2$, and
\equ(bde0) follows immediately. If H2 holds, then $aw^3 \ge -gw^2~\pp(x,y)$
and \equ(bde0) is obvious.

To prove \equ(dote0), we first assume that $(\phi_0,\phi_1) = (w(0),w_t(0))
\in Z^2_\epsilon$. In this case, one has $(w,w_t) \in C^1([0,T],Z^1_\epsilon)$,
so that $E_0 \in C^1([0,T])$, and a direct calculation shows that
$$
   \dot E_0(t) \,=\, -(1+2\epsilon c\gamma) \int_{\real^n} w_t^2 \,dz
   + \sqrt{c^2-4} \int_{\real^n} w_x w_t \,dz~. \EQ(e0dot)
$$
In the general case where $(\phi_0,\phi_1) \in Z^1_\epsilon$, we use
the fact that the solution $(w,w_t) \in Z^1_\epsilon$
depends continuously on the initial data $(\phi_0,\phi_1)$,
uniformly in $t \in [0,T]$. Therefore, if $G(t)$ denotes the right-hand
side of \equ(e0dot), we see that (for fixed $t$) both $E_0(t)$ and
$E_0(0) + \int_0^t G(s)\,ds$ are continuous functions of $(\phi_0,\phi_1)
\in Z^1_\epsilon$. Since they coincide on a dense subset (namely,
$Z^2_\epsilon$), they must be equal everywhere. This proves that $E_0
\in C^1([0,T])$ and satisfies \equ(e0dot). Finally, since
$$
   \sqrt{c^2-4} \left|\int_{\real^n} w_x w_t \,dz\right| \,\le\,
   \int_{\real^n} \left( w_t^2 + {c^2-4 \over 4} |\nabla_z w|^2 \right)\,dz~,
$$
we see that \equ(dote0) follows from \equ(bde0) and \equ(e0dot). \QED

\CLAIM Lemma(e2) Assume that H1 or H2 holds. Then there exist constants
$K_1,K_2 > 0$ depending only on $\epsilon_0,c$ such that
$$
   K_1 \|W(t)\|_{Y_\epsilon}^2 \,\le\, E_2(t) \,\le\,
   K_2 \|W(t)\|_{Y_\epsilon}^2 (1+\|w(t)\|_{X^0})~, \EQ(bde2)
$$
for all $t \in [0,T]$. Moreover, $E_2(t) \ge \alpha E_0(t)/2$, $E_2 \in
C^1([0,T])$ and
$$
   \dot E_2(t) \,\le\, -{1 \over 2} E_0(t)~. \EQ(dote2)
$$

\REMARK We recall that $W = (w,w_t)$.
The fact that $K_1,K_2$ are independent of $\epsilon$ will be
very important in Section~5, where the limiting case $\epsilon = 0$ is
considered. Note that the standard choice $\alpha = 2\epsilon$ in
\equ(e0) would lead to a constant $K_1$ of order $\epsilon$, see the proof
below.

\PROOF Since $\alpha \ge 2\epsilon$, we have
$$
   |\epsilon w w_t | \,\le\, {1 \over 3}w^2 + {3 \over 4}\epsilon^2 w_t^2
   \,\le\, {1 \over 3}w^2 + {3 \over 8}\alpha\epsilon w_t^2~. \EQ(wwt)
$$
Therefore, using \equ(bde0), we find
$$
   E_2(t) \,\ge\, \int_{\real^n} \left({\alpha \epsilon \over 8} w_t^2
   + {\alpha \over 4}|\nabla_z w|^2 + {\alpha \over 2}gw^2 + {1 \over 6}
   w^2 \right)\,dz~.
$$
Furthermore, using \equ(aw3), we obtain
$$ \eqalign{
   E_2(t) \,&\le\, \int_{\real^n} \left( \alpha\epsilon w_t^2 +
     {\alpha \over 2}|\nabla_z w|^2 + \alpha g w^2 + (1+\epsilon c\gamma)w^2
     + {\alpha \over 3}aw^3 \right)\,dz\cr
   \,&\le\, (1+K_S\|w\|_{X^0}) \int_{\real^n} \left( \alpha\epsilon w_t^2 +
     {\alpha \over 2}|\nabla_z w|^2 + \alpha g w^2 + (1+\epsilon c\gamma)w^2
     \right)\,dz~.}
$$
Since $0 \le g \le 1$, $\epsilon \le \epsilon_0$, and $1/(2c^2) \le
\alpha \le \max(2\epsilon_0,1/(2c^2))$, we arrive at \equ(bde2), with
$K_1,K_2$ independent of $\epsilon$. Similarly, since $|\epsilon w w_t|
\le w^2/2 + \alpha \epsilon w_t^2/4$, it follows from \equ(e0), \equ(bde0)
that
$$
   E_1(t) \,\ge\, -{\alpha \epsilon \over 4} \int_{\real^n} w_t^2\,dz
   \,\ge\, -{\alpha \over 2} E_0(t)~,
$$
hence $E_2(t) \ge \alpha E_0(t)/2$.

To prove \equ(dote2), we proceed along the same lines as in the preceding
lemma. Using a direct calculation and a density argument, we show that
$$
   \dot E_1(t) \,=\, \int_{\real^n} \left(\epsilon w_t^2 - 2\epsilon w_x w_t
   -|\nabla_z w|^2 -2gw^2 -aw^3 \right)\,dz~,
$$
hence
$$ \eqalign{
   \dot E_2(t) \,&=\, (\epsilon - \alpha(1+2\epsilon c\gamma))\int_{\real^n}
     w_t^2\,dz + (\alpha\sqrt{c^2-4} - 2\epsilon c) \int_{\real^n} w_x w_t
     \,dz \cr
   \,&-\, \int_{\real^n} \left(|\nabla_z w|^2 + 2gw^2 + aw^3 \right)\,dz~.}
$$
If $\alpha = 2\epsilon$, then $\alpha\sqrt{c^2-4} -2\epsilon c = -4\epsilon
\gamma$, and
$$
   4\epsilon \gamma|w_x w_t| \,\le\, 4\epsilon^2 c \gamma w_t^2 + {\gamma
   \over c} w_x^2 \,\le\, 2\epsilon\alpha c\gamma w_t^2 + {1 \over 2}
   |\nabla_z w|^2~,
$$
since $\gamma/c \le 1/2$ by \equ(gamma). If $\alpha = 1/(2c^2) \ge 2\epsilon$,
then $|\alpha\sqrt{c^2-4}-2\epsilon c| \le \alpha c$ and
$$
   \alpha c |w_x w_t| \,\le\, {\alpha \over 4} w_t^2 + \alpha c^2 w_x^2
   \,=\, {\alpha \over 4} w_t^2 + {1 \over 2} |\nabla_z w|^2~.
$$
In both cases, we find
$$
   \dot E_2(t) \,\le\, -\int_{\real^n} \left( {\epsilon \over 2} w_t^2
   + {1 \over 2} |\nabla_z w|^2 + 2gw^2 + aw^3 \right)\,dz.
$$

If H1 holds, then by \equ(e0), \equ(aw3)
$$ \eqalign{
   \dot E_2(t) + {1 \over 2}E_0(t) \,&\le\, -\int_{\real^n}
     \left( {1 \over 4}|\nabla_z w|^2 + {3 \over 2}gw^2 + {5 \over 6}
     aw^3 \right)\,dz \cr
   \,&\le\, -(1-5K_S\|w\|_{X^0}) \int_{\real^n} \left( {1 \over 4}
     |\nabla_z w|^2 + {1 \over 2}gw^2 \right)\,dz \,\le\, 0~.}
$$
If H2 holds, then we simply have
$$
   \dot E_2(t) + E_0(t) \,\le\, -\int_{\real^n} \left(gw^2 + {2 \over 3}
   aw^3 \right)\,dz \,\le\, -{1 \over 3}\int_{\real^n} gw^2\,dz \,\le\, 0~.
$$
In both cases, we obtain \equ(dote2). \QED

\REMARK Up to now, we did not use the fact that $d > 0$. Therefore, \clm(e0)
and \clm(e2) are still valid if H2 holds with $d = 0$, and the constants
$K_1,K_2$ are independent of $d$.

\SUBSECTION Weighted Functionals

Under the same assumptions as above, we define the weighted functionals
$$ \eqalign{
   F_0(t) \,&=\, \int_{\real^n} \left({\epsilon \over 2} a^2 w_t^2
     + {1 \over 2} a^2|\nabla_z w|^2 + a^2 g w^2 + {1 \over 3} a^3 w^3
     \right) \,dz ~,\cr
   F_1(t) \,&=\, \int_{\real^n} \left({1-2\epsilon c\gamma \over 2}a^2 w^2
     + \epsilon a^2 w w_t \right) \,dz~,\cr
   F_2(t) \,&=\, \ha F_0(t) + F_1(t) + \beta E_0(t)~,\phantom{1 \over 2}}
   \EQ(f2)
$$
where $\ha = \max(2\epsilon,d/(2c^2))$ and $\beta = 3\ha$. In the case
where H1 holds, we set $d =1$, so that $\ha = \alpha$.

\REMARK The additional term $\beta E_0(t)$ in \equ(f2) guarantees that
$F_2(t) \ge 0$. However, if $\epsilon$ is sufficiently small, then
$\ha F_0(t) + F_1(t)$ is already positive, so we may set $\beta = 0$. This
possibility will be used in Section~4.3 below.

\CLAIM Lemma(f2) Assume that H1 or H2 holds. Then there exist constants
$K_3,K_4,K_5 > 0$ such that
$$
   K_3 \|W(t)\|_{Y_{\epsilon a}}^2 \,\le\, F_2(t) \,\le\, K_4
   \|W(t)\|_{Y_{\epsilon a}}^2 (1+\|w(t)\|_{X^0}) + \beta E_0(t)~, \EQ(bdf2)
$$
for all $t \in [0,T]$. Moreover, $F_2 \in C^1([0,T])$ and satisfies
$$
   \dot F_2(t) + \kappa F_2(t) \,\le\, K_5 E_0(t)~, \EQ(dotf2)
$$
where $\kappa = d/(8(1+\ha))$.

\REMARK Here and in the sequel, $K_3,K_4,\dots$ denote positive constants
depending only on $\epsilon_0,c$ and, if H2 holds, on $d > 0$.

\PROOF Using the identity
$$
   \int_{\real^n} a^2 |\nabla_z w|^2 \,dz \,=\, \int_{\real^n}
   \left(|\nabla_z(aw)|^2 + \gamma^2 a^2 w^2 \right)\,dz~, \EQ(waw)
$$
together with the relation $1 - c\gamma + \gamma^2 = 0$, we write $F_2(t)$
as
$$ \eqalign{
   F_2(t) \,=\, \int_{\real^n} &\left({\ha \epsilon \over 2} a^2 w_t^2
     + {\ha \over 2} |\nabla_z(aw)|^2 + c\gamma(\ha/2-\epsilon)a^2 w^2
     + {\ha \over 2} (2g-1) a^2 w^2 \right.\cr
   \,&+\, \left.{\ha \over 3} a^3 w^3 + {1 \over 2} a^2 w^2
     + \epsilon a^2 w w_t \right)\,dz + \beta E_0(t)~.} \EQ(newf2)
$$
To prove \equ(bdf2), we first note that $c\gamma(\ha/2-\epsilon) \le
c\gamma \ha/2 \le \ha$ and $g\le 1$. Using \equ(wwt), we thus find
$$
   F_2(t) \,\le\, \int_{\real^n} \left(\ha \epsilon a^2 w_t^2 +
     {\ha \over 2} |\nabla_z(aw)|^2 + (1+\ha) a^2 w^2 + {\ha \over 3}
     a^3 w^3 \right)\,dz + \beta E_0(t)~. \EQ(upf2)
$$
Furthermore, in analogy with \equ(xneg), we have
$$
   \left| \int_{\real^n} a^3 w^3 \,dz \right| \,\le\, K_S \|w\|_{X^0}
   \int_{\real^n} \left( a^2 w^2 + |\nabla_z(aw)|^2 \right)\,dz~.
   \EQ(a3w3)
$$
Therefore, combining \equ(upf2) and \equ(a3w3), we easily obtain the
upper bound in \equ(bdf2).

To prove the lower bound, we first use \equ(wwt)
and the fact that $\ha \ge 2\epsilon$. We find
$$ \eqalign{
   F_2(t) \,\ge\, \int_{\real^n} &\left( {\ha \epsilon \over 8} a^2 w_t^2
     + {\ha \over 2} |\nabla_z(aw)|^2 + {\ha \over 2}(2g-1) a^2 w^2
     + {\ha \over 3}a^3 w^3 + {1 \over 6} a^2 w^2 \right)\,dz \cr
     \,&+\, \beta E_0(t)~.} \EQ(midf2)
$$
If H1 holds, we apply \equ(a3w3). Since $K_S \|w\|_{X^0} \le 3/4$, we obtain
$$
   F_2(t) \,\ge\, \int_{\real^n} \left({\ha \over 8} a^2 w_t^2 +
   {\ha \over 4}|\nabla_z(aw)|^2 + \ha(g-3/4)a^2 w^2 + {1 \over 6}a^2 w^2
   \right)\,dz + \beta E_0(t)~.
$$
If $x \le 0$, then $g(x)-3/4 \ge 1-\sigma-3/4 \ge 0$, since $\sigma \le 1/8$.
If $x \ge 0$, then $g(x)-3/4 \ge -1$ and $a(x)^2 \le a(x) \le 3g(x)/2$, so
that
$$
   \int_{x\ge 0} a^2 w^2 \,dz \,\le\, {3 \over 2} \int_{x\ge 0} g w^2 \,dz
   \,\le\, 3 E_0(t)~, \EQ(xpose0)
$$
by \equ(bde0). Therefore, since $\beta = 3\ha$, we have
$$
   F_2(t) \,\ge\, \int_{\real^n} \left({\ha \epsilon \over 8} a^2 w_t^2 +
   {\ha \over 4}|\nabla_z(aw)|^2 + {1 \over 6} a^2 w^2 \right)\,dz~.
   \EQ(lowf2)
$$
If H2 holds, we observe that
$$
   \ha(g-1/2)a^2 w^2 + {\ha \over 3}a^3w^3 \,\ge\, \ha (2g/3 -1/2)a^2 w^2
   \quad \pp(x,y)~.
$$
Again, we have $2g(x)/3 -1/2 \ge (1-4\sigma)/6 \ge 0$ if $x \le 0$, and
$2g(x)/3 - 1/2 \ge -1$ if $x \ge 0$. Therefore, using \equ(midf2)
and proceeding as above, we again arrive at \equ(lowf2). This proves
the lower bound in \equ(bdf2).

To prove \equ(dotf2), we proceed along the same lines as in the preceding
lemmas. Using a direct calculation and a density argument, we first show
that
$$ \eqalign{
   \dot F_2(t) \,&=\, (\epsilon-\ha) \int_{\real^n} a^2 w_t^2 \,dz +
     c(\ha-2\epsilon) \int_{\real^n} a^2 w_x w_t \,dz \cr
   \,&-\, \int_{\real^n} \left(|\nabla_z(aw)|^2 + (2g-1)a^2 w^2 + a^3 w^3
     \right)\,dz + \beta \dot E_0(t)~.} \EQ(f2dot)
$$
If $\ha = d/(2c^2) > 2\epsilon$, then
$$
   |c(\ha - 2\epsilon)w_x w_t| \,\le\, c\ha|w_x w_t| \,\le\, {\ha \over 4}
   w_t^2 + {d \over 2} |\nabla_z w|^2~.
$$
Therefore, using \equ(waw) and the fact that $\gamma^2 \le 1$, we find
$$
   \dot F_2(t) \,\le\, - \int_{\real^n} \left({\epsilon \over 2} a^2 w_t^2
   + {1 \over 2} |\nabla_z(aw)|^2 + (2g-1-d/2)a^2 w^2 + a^3 w^3 \right)\,dz
   + \beta \dot E_0(t)~. \EQ(lastf2dot)
$$
If $\ha = 2\epsilon$, then \equ(lastf2dot) follows immediately from
\equ(f2dot).

We now combine \equ(upf2) and \equ(lastf2dot). Using \equ(dote0) and
the fact that $\kappa \ha \le 1/2$, we easily find
$$ \eqalign{
   \dot F_2(t) + \kappa F_2(t) \,&\le\, - \int_{\real^n} \left( {1 \over 4}
     |\nabla_z(aw)|^2 + (2g-1-d/2-\kappa(1+\ha))a^2 w^2 \right)\,dz \cr
   \,&-\, (1-\kappa\ha/3) \int_{\real^n} a^3 w^3 \,dz + \hb E_0(t)~,}
   \EQ(f2last)
$$
with $\hb = \beta(\kappa+c^2-4)$. If H1 holds, we use \equ(a3w3) and obtain
$$
   \dot F_2(t) + \kappa F_2(t) \,\le\, - \int_{\real^n} \bigl(2g-3/2
   -\kappa(1+\ha) - K_S \|w\|_{X^0}\bigr) a^2 w^2 \,dz + \hb E_0(t)~.
$$
If $x \le 0$, then $2g(x)-3/2-\kappa(1+\ha)-K_S\|w\|_{X^0} \ge 1/2-2\sigma
-\kappa(1+\ha) -K_S \delta \ge 0$, by assumptions on $\sigma,\delta,\kappa$.
If $x \ge 0$, the same quantity is bounded from below by $-2$. Therefore,
using \equ(xpose0), we find $\dot F_2 + \kappa F_2 \le (6+\hb)E_0$,
which is \equ(dotf2).

If H2 holds, we infer from \equ(f2last)
$$
   \dot F_2(t) + \kappa F_2(t) \,\le\, -\int_{\real^n} \bigl((1+d)g - 1
   -d/2 -\kappa(1+\ha)\bigr) a^2 w^2 \,dz + \hb E_0(t)~.
$$
Since $g(x) \to 1$ as $x \to -\infty$, there exists $x_d \le 0$
such that $g(x) \ge 1-d/8$ for all $x \le x_d$. Therefore, if $x \le x_d$,
one has $(1+d)g-1-d/2-\kappa(1+\ha) \ge (3d-d^2)/8 -\kappa(1+\ha) \ge 0$
by assumptions on $d,\kappa$. If $x \ge x_d$, the same quantity
is bounded from below by $-2$, and
$$
   \int_{x \ge x_d} a^2 w^2 \,dz \,\le\, {9 \over 4} \,e^{-2\gamma x_d}
   \int_{x \ge x_d} g w^2 \,dz \,\le\, {9 \over 2}\,e^{-2\gamma x_d} E_0(t)~,
$$
since $3g(x) \ge 3g(x-x_d) \ge 2a(x-x_d) = 2e^{\gamma x_d}a(x)$ for all
$x \ge x_d$. Combining these inequalities, we find $\dot F_2 + \kappa F_2 \le
(9 e^{-2\gamma x_d}+\hb) E_0$, which is the desired result. \QED

\CLAIM Corollary(wbound) Assume that H1 or H2 holds. Then there exists a
constant $K_6 \ge 1$ such that
$$
   \|W(t)\|_{Z^1_\epsilon} \,\le\, K_6 \|W(0)\|_{Z^1_\epsilon}
   (1+\|w(0)\|_{X^0})^{1/2}~, \EQ(wbound)
$$
for all $t \in [0,T]$.

\PROOF According to \clm(e2), we have
$$
   \|W(t)\|_{Y_\epsilon}^2 \,\le\, {K_2 \over K_1} \|W(0)\|_{Y_\epsilon}^2
   (1+\|w(0)\|_{X^0})~, \EQ(one)
$$
for all $t \in [0,T]$, since $E_2$ is a decreasing function of $t$.
On the other hand, it follows from \equ(dotf2) and \clm(e2) that
$\dot F_2(t) + \kappa F_2(t) \le \hat K_5 E_2(t)$, where $\hat K_5 =
2 K_5/\alpha$. Integrating this inequality, we find
$$
   F_2(t) \,\le\, e^{-\kappa t} F_2(0) + \hat K_5 \int_0^t e^{-\kappa (t-\tau)}
   E_2(\tau)\,d\tau \,\le\, F_2(0) + {\hat K_5 \over \kappa} E_2(0)~,
$$
hence
$$
   \|W(t)\|_{Y_{\epsilon a}}^2 \,\le\, {1 \over K_3} \left( K_4
   \|W(0)\|_{Y_{\epsilon a}}^2 + (6 + \hat K_5/\kappa)K_2
   \|W(0)\|_{Y_\epsilon}^2 \right) (1+\|w(0)\|_{X^0})~, \EQ(two)
$$
by \equ(bde2), \equ(bdf2). Combining \equ(one) and \equ(two),
the result follows. \QED

If H2 holds with $d=0$, it is no longer possible to bound $W(t)$ uniformly
in time as in \clm(wbound), but the energy estimates above still imply
that $\|W(t)\|_{Z^1_\epsilon}$ cannot grow faster than an exponential.
This result will be useful in Section~4.

\CLAIM Corollary(d=zero) Assume that H2 holds with $d=0$. Then there
exist constants $\rho > 0$ and $K_7 \ge 1$ such that
$$
   \|W(t)\|_{Z^1_\epsilon} \,\le\, K_7(1+e^{\rho t})\|W(0)\|_{Z^1_\epsilon}
   (1+\|w(0)\|_{X^0})^{1/2}~, \EQ(wd0)
$$
for all $t \in [0,T]$.

\PROOF We recall that \clm(e0) and \clm(e2) still hold if $d = 0$,
see the remark at the end of Section~2.1. Furthermore, if we define $F_2(t)$
by \equ(f2) with $\ha = \alpha = \max(2\epsilon,1/(2c^2))$ and $\beta =
3\alpha$, then it is easily verified that \equ(bdf2) is still valid.
However, \equ(lastf2dot) has to be replaced by
$$
   \dot F_2(t) \,\le\, - \int_{\real^n} \left({\epsilon \over 2} a^2 w_t^2
   + {1 \over 2} |\nabla_z(aw)|^2 + (g-3/2) a^2 w^2 \right)\,dz +
   \beta \dot E_0(t)~.
$$
Therefore, using \equ(dote0), \equ(lowf2), we obtain
$$
   \dot F_2(t) \,\le\, 9 F_2(t) + \beta(c^2-4) E_0(t)~,
$$
which replaces \equ(dotf2). Integrating this inequality and proceeding
as in the proof of \clm(wbound), we obtain \equ(wd0), with
$\rho = 9/2$. \QED

\SECTION Local Stability

In this section, we prove \clm(local) using the energy estimates of
Section~2.

\medskip\noindent{\bf Proof of \clm(local).}
Let $\delta_0, K_0$ be defined by the relations
$$
   K_6 \delta_0 (1+\delta_0)^{1/2} \,=\, \delta/2~, \quad
   K_0 \,=\, K_6 (1+\delta_0)^{1/2}~,
$$
where $K_6 \ge 1$ is given in \clm(wbound) and $\delta = 1/(8K_S)$ in the
assumption H1, Section~2. Then, for all $\Phi = (\phi_0,\phi_1) \in
Z^1_\epsilon$ such that $\|\Phi\|_{Z^1_\epsilon} \le \delta_0$, Eq.\equ(w)
has a unique global solution $W(t) = (w(t),w_t(t)) \in Z^1_\epsilon$
satisfying $W(0) = \Phi$. Indeed, in view of the local existence result
(\clm(loc-ex)), it is sufficient to show that $\|W(t)\|_{Z^1_\epsilon} <
\delta$ whenever $W(t)$ exists. Assume on the contrary that there exists
a time $T > 0$ such that $\|W(T)\|_{Z^1_\epsilon} = \delta$ and
$\|W(t)\|_{Z^1_\epsilon} < \delta$ for all $t \in [0,T)$. Then
$\|w(t)\|_{X^0} \le \|W(t)\|_{Z^1_\epsilon} \le \delta$ for all $t \in
[0,T]$, so that H1 holds on $[0,T]$. By \clm(wbound), it follows that
$\|W(T)\|_{Z^1_\epsilon} \le K_6 \delta_0(1+\delta_0)^{1/2} = \delta/2$,
which is a contradiction. Therefore, $W(t)$ exists for all times and the
assumption H1 is always satisfied. By \clm(wbound) again, we conclude that
$\|W(t)\|_{Z^1_\epsilon} \le K_0 \|\Phi\|_{Z^1_\epsilon}$ for all $t \ge 0$,
which proves \equ(wwtbd).

To prove \equ(tozero), \equ(decay), we use the differential inequalities
satisfied by the functionals $E_0, E_2, F_2$ defined in Section~2.
The following arguments are standard (see for example [EW]) and will
be reproduced here for the sake of completeness. First, since $E_2$
is a positive, decreasing function of $t$ by \clm(e2), $E_2(t)$ converges
to a nonnegative limit as $t \to +\infty$. By \equ(dote0), \equ(dote2),
so does $E_0 + 2(c^2-4)E_2$. Therefore, $E_0(t)$ converges as $t \to
+\infty$, and since $E_0(t) \ge 0$ it follows from \equ(dote2) that the
integral
$$
   \int_0^{+\infty} E_0(\tau)\,d\tau \,\le\, 2(E_2(0)-E_2(+\infty))
$$
is finite. Thus $E_0(t)$ converges to zero as $t \to +\infty$. Moreover,
integrating the differential inequality \equ(dotf2), we find
$$ \eqalign{
   F_2(t) \,&\le\, e^{-\kappa t} F_2(0) + K_5 \int_0^t e^{-\kappa (t-\tau)}
     E_0(\tau)\,d\tau \cr
   \,&\le\, e^{-\kappa t}F_2(0) + K_5 \left(e^{-\kappa t/2} \int_0^{t/2}
     E_0(\tau) \,d\tau + \int_{t/2}^t e^{-\kappa(t-\tau)} E_0(\tau)\,d\tau
     \right)\cr
   \,&\le\, e^{-\kappa t/2} (F_2(0) + 2K_5 E_2(0)) + {K_5 \over \kappa}
     \sup_{\tau \in [t/2,t]} E_0(\tau)~,} \EQ(gron)
$$
hence $F_2(t)$ converges to zero as $t \to +\infty$. Therefore, using the
lower bounds in \equ(bde0), \equ(bdf2), we obtain \equ(tozero).

In the case where $c=2$, $E_0$ itself is a decreasing function of $t$
by \equ(dote0), hence $tE_0(t) \le 2 \int_{t/2}^t E_0(\tau)d\tau$.
Thus $tE_0(t)$ converges to zero as $t \to +\infty$, and
by \equ(gron) the same is true for $tF_2(t)$. Therefore, using again
\equ(bde0), \equ(bdf2), we obtain \equ(decay). This concludes the proof of
\clm(local). \QED

\SECTION Global Stability Results in the One-Dimensional Case

Throughout this section, we assume that $n=1$. First, we prove \clm(global)
using the results of Section~2 and the Maximum Principle for hyperbolic
equations. Then, we study in more details the case $\epsilon \le 1/4$;
we give linear upper and lower bounds for the solutions of \equ(w).
Finally, using these linear bounds, we prove \clm(decay).

\SUBSECTION Global Stability in the General Case

We first show that the assumption H2 (see Section~2) holds if the solution
$w(x,t)$ of \equ(w) is bounded from above and if the initial data satisfy
\equ(phi0), \equ(phi1).

\CLAIM Proposition(41) Let $\epsilon > 0$, $d \in [0,1]$ and let $K$ be a
nonnegative constant such that
$$
   1- 4\epsilon (d+K) \ge 0~. \EQ(41)
$$
For some $T > 0$, assume that $(w,w_t) \in C^0([0,T],Z^1_\epsilon)$
is a solution of \equ(w) with initial data $(\phi_0,\phi_1)$ satisfying
\equ(phi0), \equ(phi1), namely
$$
  \phi_0(x) \,\ge\, -(1 - d)a^{-1}(x)g(x)~,
  \EQ(42)
$$
$$
  \epsilon \phi_1(x) \,\ge\, \epsilon c \phi_0'(x) - {1\over 2}
  \bigl(\phi_0 + (1-d)a^{-1}g\bigr)(x) + \epsilon c \bigl( -\gamma\phi_0
  + (1-d)a^{-1}g'\bigr)(x)~, \EQ(43)
$$
for (almost) every $x \in \real$. Suppose moreover that
$$
   a(x) w(x,t) \le K~, \quad \forall\, (x,t) \in \real \times [0,T]~. \EQ(44)
$$
Then
$$
  w(x,t) \ge  -(1 - d)a^{-1}(x)g(x)~,\quad \forall\, (x,t) \in \real
  \times [0,T]~. \EQ(45)
$$

\PROOF We recall that the inequality \equ(45) is equivalent to $v(x,t) \ge
dg(x)$, where $v(x,t)$ is the solution of \equ(v). Also, we remark that
$dg - v$ belongs to the space $C^0([0,T], H_{loc}^1(\real)) \cap C^1([0,T],
L_{loc}^2(\real))$ and satisfies
$$
  \tilde L(dg-v) \, \equiv \, \tilde \LL(dg-v) + \tilde h(x,t) (dg-v)
  \, = \, g^2 d (1 - d )~,
$$
where $\tilde \LL(v) \, =\, v_{xx} + 2\epsilon c v_{xt} - \epsilon v_{tt}
+ cv_x -v_t$ and $\tilde h(x,t) \, = \, 1 - v(x,t) - dg(x)$.
Therefore, to prove \equ(45), we are led to apply the Maximum Principle
(Theorem~A.1, Appendix~A) to the function $dg-v$ and to the operator
$\tilde L$. The condition (A.1) is obviously satisfied. Due to
\equ(44), the condition (A.2) holds, i.e.,
$$
  \tilde h(x,t) \ge 1-(1+d)g(x) - K \ge -K - d ~.
$$
This estimate and \equ(41) imply that (A.3) also holds. Moreover, since
$0 \le d \le 1$, we have $\tilde L(dg - v)(x,t) \ge 0~\pp(x,t) \in  \real
\times [0,T]$, which is (A.4). Finally, the conditions (A.5) and (A.6)
required on $dg - v$ are nothing else but the hypotheses \equ(42) and
\equ(43). Therefore, it follows from Theorem~A.1 that $(dg -v)(x,t) \le 0$,
that is, $w(x,t) \ge -(1-d)a(x)^{-1}g(x)$, for all $(x,t) \in \real \times
[0,T]$. \QED

Using \clm(41) and \clm(wbound), we now prove the first global stability
result.

\medskip\noindent{\bf Proof of \clm(global).} The proof is very similar to
the one  of \clm(local) given in Section~3. Let $\mu$ be a real number,
$0 < \mu < 1$. We define $K^*$ by the relation
$$
  K_6 K^*(1 + K^*)^{1/2} \, =\, (1-\mu)K~,
$$
where $K_6 \ge 1$ has been introduced in \clm(wbound). According to
\clm(loc-ex), there exist a time $T > 0$ and a unique solution $W(t)=(w,w_t)
\in C^0([0,T), Z_\epsilon^1)$ of \equ(w) with initial data $(\phi_0,\phi_1)$
such that $\| W(t) \|_{Z_\epsilon^1} <  K$ for all $t \in [0,T)$ and, if
$T < \infty $, $W(t) \in C^0([0,T],Z^1_\epsilon)$ and
$\| W(T) \|_{Z_\epsilon^1} = K$.
We show by contradiction that $T= \infty$. If $T < \infty$, we have
$$
   a(x)w(x,t) \,\le\, \|aw(t)\|_{L^\infty} \,\le\, \|aw(t)\|_{H^1} \,\le\,
   \|W(t)\|_{Z^1_\epsilon} \,\le\, K~,
$$
for all $t \in [0, T]$. Thus, by \clm(41), $w(x,t) \ge -(1-d)a(x)^{-1}g(x)~$
for all $(x,t) \in \real \times [0,T]$, i.e. the assumption H2 of Section~2
holds on $[0,T]$ (we recall that $d > 0$ here). By \clm(wbound), it follows
that $\| W(T) \|_{Z_\epsilon^1} \le  K_6( 1+ K^*)^{1/2} K^* = (1 - \mu)K$,
which is a contradiction. Therefore $T = \infty$, and the inequalities
\equ(lwbd) hold for all times. The properties \equ(tozero),
\equ(decay) are proved like in Section 3. \QED

\SUBSECTION Linear Bounds in the Case $\epsilon \le 1/4$

{}From now on, we assume that $\epsilon \le 1/4$. In this case, the range
of application of the Maximum Principle is much wider, and we can show that
the solution $w(x,t)$ of \equ(w) is bounded from above and from below by
solutions of suitable linear equations. These linear bounds will be crucial
for the proof of \clm(decay). Before stating the results, we introduce some
additional notation.

For all $d \in [-1,1]$, we denote by $S_d(t) \in
\LL(Z_\epsilon^1,Z_\epsilon^1)$ the linear group associated with the equation
$(L_d w)(x,t) = 0$, where
$$
  L_d w \,=\, -\epsilon w_{tt} - (1 + 2\epsilon c \gamma) w_t
  + 2\epsilon c w_{xt} + w_{xx} + \sqrt{c^2 - 4}\,w_x - (1+d)g w~. \EQ(ld)
$$
For $(\phi_0,\phi_1) \in Z_\epsilon^1$, we set
$$
  S_d(t)(\phi_0,\phi_1) \,=\, (\tilde w_d(t), \tilde w_{dt}(t))~. \EQ(Sd)
$$
In \equ(phi+), \equ(phi-), we have defined the positive and negative parts
$(\phi_0^\pm,
\phi_1^\pm)$ of $(\phi_0,\phi_1)$. In analogy with \equ(Sd), we set
$$
  S_d(t)(\phi_0^\pm,\phi_1^\pm) \,=\,
  (\tilde w_d^\pm(t), \tilde w_{dt}^\pm(t))~. \EQ(Sd+)
$$

We now show the existence of a linear upper bound.

\CLAIM Lemma(43) Let $\epsilon \le 1/4$. For any $(\phi_0, \phi_1)
\in Z_\epsilon^1$, the solution $(w,w_t) \in C^0([0,T],Z^1_\epsilon)$
of \equ(w), with initial data $(\phi_0, \phi_1)$, satisfies
for any $d \in [-1,1]$,
$$
  w(x,t) \,\le\, \tilde w_1(x,t) \,\le\, \tilde w_d^+(x,t)~, \quad \forall\,
  (x,t) \in \real \times [0,T]~, \EQ(412)
$$
where $\tilde w_1$, $\tilde w_d^+$ have been defined in \equ(Sd) and \equ(Sd+)
respectively.

\PROOF We first prove the inequality $w(x,t) \le \tilde w_1(x,t)$.
The function $w - \tilde w_1$ satisfies the equation $L_1(w - \tilde w_1)
= aw^2 \ge 0$, where $L_1$ has been defined in \equ(ld). Thus, we can apply
the Maximum Principle to the function $w - \tilde w_1$ and to
the operator $L_1$. Indeed the conditions (A.1), (A.2), (A.4) are satisfied,
and, since the initial data for $w,\tilde w$ coincide, (A.5) and (A.6)
obviously hold. Since $-2 g(x) \ge -2$, the condition (A.3) with $\underline
h = -2$ becomes $(1-4 \epsilon) (\epsilon + \epsilon^2 c^2) \,\ge\, 0$,
which is satisfied because $\epsilon \le 1/4$. Therefore, Theorem~A.1.
implies that $w(x,t) - \tilde w_1(x,t) \le 0$ for all $(x,t) \in \real
\times [0,T]$.

We next show that $\tilde w_d^+(x,t) \ge 0$. Since $L_d (-\tilde w_d^+) = 0$,
we can apply the Maximum Principle to the function $-\tilde w_d^+$ and to the
operator $L_d$. In view of the first part of the proof, the conditions (A.1)
to (A.4) hold. Due to the choice of $(\phi_0^+, \phi_1^+)$ made in
\equ(phi+), the hypotheses (A.5) and (A.6) are also satisfied. Therefore,
by Theorem~A.1, $\tilde w_d^+(x,t) \ge 0$ for all $(x,t) \in \real \times
\real_+$.

Finally, we show that $\tilde w_1(x,t) \le \tilde w_d^+(x,t)$ for all
$d \in [-1,1]$, by applying the Maximum Principle to the function
$\tilde w_1 - \tilde w_d^+$ and to the operator $L_1$. As we have already
remarked, the hypotheses (A.1), (A.2), (A.3) are satisfied. The condition
(A.4) holds, since $L_1(\tilde w_1 - \tilde w_d^+) = (1 - d) g
\tilde w_d^+$ and $\tilde w_d^+ \ge 0$. The choice of
$(\phi_0^+, \phi_1^+)$ in \equ(phi+) also implies that (A.5) and (A.6) hold.
Hence, by Theorem~A.1, $\tilde w_1(x,t) \le \tilde w_d^+(x,t)$ for all
$(x,t) \in \real \times [0,T]$. \QED

In a similar way, we obtain linear lower bounds for $w(x,t)$.

\CLAIM Lemma(44) Let $\epsilon \le 1/4$, $d \in [0,1]$, and let $K$ be a
nonnegative constant such that
$$
  1- 4\epsilon (1+K) \,\ge\, 0~. \EQ(422)
$$
For some $T > 0$, assume that $(w,w_t) \in C^0([0,T],Z^1_\epsilon)$ is
a solution of \equ(w) with initial data $(\phi_0, \phi_1) \in
Z_\epsilon^1$ satisfying \equ(42) and \equ(43). Suppose moreover that \equ(44)
holds. Then
$$
\tilde w_{-1}^-(x,t) \le \tilde w_d^-(x,t) \le  w(x,t)~, \quad \forall\,
(x,t) \in \real \times [0,T]~, \EQ(423)
$$
where $\tilde w_{-1}^-$, $\tilde w_d^-$ have been defined in \equ(Sd+).

\PROOF
As in the proof of \clm(43), we show that $\tilde w_d^-(x,t) \le 0$ by
applying the Maximum Principle to the function $\tilde w_d^-$ and to
the operator $L_d$.
To show that $\tilde w_{-1}^-(x,t) \le \tilde w_d^-(x,t)$, we apply the
Maximum Principle to the function $\tilde w_{-1}^- - \tilde w_d^-$ and
to the operator $L_{-1}$. Since $L_{-1} (\tilde w_{-1}^- - \tilde w_d^-)
= -(1 + d)g \tilde w_d^-$ and $\tilde w_d^- \le 0$, the hypothesis (A.4)
holds. The other conditions are obvious or have been verified in the proof
of \clm(43).

It remains to prove that $w(x,t) \ge \tilde w_d^-(x,t)$. We again apply the
Maximum Principle, but now to the function $\tilde w_d^- - w$ and to the
operator $L_1^* = L_{-1} + h^*$, where $h^*(x,t)\,=\, -(2g(x) + a(x)w(x,t))$.
Since $h^*(x,t) \ge -2 -K$, the condition (A.3) becomes $(1 - 4 \epsilon
(1 + K)) (\epsilon + \epsilon^2 c^2) \ge 0$, which is nothing but
\equ(422). Moreover, we have $L_1^*(\tilde w_d^- - w) = -\tilde w_d^-
(aw + (1-d)g) \ge 0$, since $\tilde w_d^-(x,t) \le 0$ and $a(x)w(x,t)
\ge -(1-d)g(x)$ by \clm(41). Thus (A.4) holds, and due to the choice of
$(\phi_0^-, \phi_1^-)$ in \equ(phi-) the conditions (A.5) and (A.6) are
also satisfied. Therefore $\tilde w_d^-(x,t) \le w(x,t)$ by Theorem~A.1,
and \clm(44) is proved. \QED

Since $\tilde w_1(x,t)$ is a solution of the linear equation $L_1 w = 0$, it
is easy to bound it in terms of the initial data $(\phi_0,\phi_1)$.
We have the following result:

\CLAIM Lemma(45) Let $\epsilon \le 1/4$. There exists a constant
$N = N(c) \ge 1$ such that
$$
   \|S_1(t)\|_{\LL(Z^1_\epsilon,Z^1_\epsilon)} \,\le\, N~, \quad
   \forall \,t \in \real_+~. \EQ(sdbd)
$$

\PROOF All we need is repeating the energy estimates of Section~2 for
the linear equation obtained by dropping the last term $-aw^2$ in \equ(w).
The functionals $E_0(t), E_2(t), F_2(t)$ are then replaced by the quadratic
expressions
$$ \eqalign{
   \tilde E_0(t) \,&=\, \int_{\real} \left({\epsilon\over 2} w_t^2 +
     {1 \over 2} |w_x|^2 + g w^2 \right)\,dx~, \quad
     \tilde E_2(t) \,=\, \alpha \tilde E_0(t) + E_1(t)~,\cr
   \tilde F_2(t) \,&=\, \alpha\int_{\real} \left({\epsilon\over 2} a^2 w_t^2 +
     {1 \over 2} a^2 |w_x|^2 + a^2 g w^2 \right) dx + F_1(t)
     + \beta \tilde E_0(t)~,}
$$
where $\alpha,\beta,E_1(t),F_1(t)$ are defined in \equ(e0), \equ(f2). Of
course, the assumptions H1, H2 are not needed anymore, since they were used
to control the cubic terms in $E_2(t), F_2(t)$. Following exactly
the lines of the proof of \clm(e0), \clm(e2), \clm(f2) (with obvious
simplifications), we arrive at \clm(wbound), which reduces in this case to
$\|W(t)\|_{Z^1_\epsilon} \le N \|W(0)\|_{Z^1_\epsilon}$ for some
$N(c) \ge 1$. This proves \equ(sdbd). \QED

\SUBSECTION Global Stability and Decay in the Case $\epsilon \le 1/4$

Using the linear bounds of the previous paragraph, we are now able to improve
the global stability results. \clm(decay) will be a direct consequence of the
following two propositions:

\CLAIM Proposition(46) Let $\epsilon \le 1/4$, $d \in [0,1]$
and $K$ be a nonnegative constant, such that \equ(422) holds. For any
$(\phi_0, \phi_1) \in Z_{\epsilon}^1$ satisfying \equ(42), \equ(43) and
$$
  \inf (\|(\phi_0,\phi_1)\|_{Z_\epsilon^1},
  \|(\phi_0^+,\phi_1^+)\|_{Z_\epsilon^1}) \,\le\, {K \over N}~, \EQ(433)
$$
where $(\phi_0^+,\phi_1^+)$ is defined in \equ(phi+) and
$N$ in \clm(45), there exists a unique global solution $(w,w_t) \in
C^0(\real_+, Z_{\epsilon}^1)$ of \equ(w) with initial data
$(\phi_0, \phi_1)$. Moreover, we have
$$
   -(1-d)g(x) \,\le\, a(x)w(x,t) \,\le\, K~,\quad \forall\, (x,t) \in
   \real \times \real_+~, \EQ(434)
$$
and
$$
   \tilde w_{-1}^-(x,t) \,\le\,  w(x,t) \,\le\,
   \tilde w_1(x,t) \,\le\, \tilde w_{-1}^+(x,t)~, \quad \forall\,
   (x,t) \in \real \times \real_+~. \EQ(435)
$$
In addition, if $d>0$, the properties \equ(tozero) and \equ(decay) hold.

\REMARK The case $K=0$ is non trivial, because it corresponds to
$(\phi_0^+,\phi_1^+) = 0$, i.e.
$$
   \phi_0(x) \,\,\le\,\, 0~, \quad \phi_1(x) \,\,\le\,\, c\phi_{0}'(x) -
   ({1 \over 2\epsilon}+c\gamma) \phi_0(x)~.
$$
In this case, \equ(434) shows that $w(x,t)$ stays nonpositive for all times.

\PROOF According to \clm(loc-ex), there exist  a maximal time $T > 0$ and a
solution $(w,w_t) \in C^0([0,T), Z_\epsilon^1)$ of \equ(w) with initial data
$(\phi_0,\phi_1)$ such that either $T = \infty$, or $T < \infty$. In the
latter case, there exists a sequence of positive times $t_n$, $t_n < T$, such
that $t_n\to T$ as $n\to +\infty$ and
$$
  \|(w(t_n),w_t(t_n))\|_{Z_\epsilon^1} \to +\infty~, \EQ(436)
$$
as $n \to +\infty$. By \clm(43) and \clm(45), we have
$$ \eqalign{
   a(x) w(x,t) \,&\le\, \inf(a(x) \tilde w_1(x,t), a(x) \tilde w_1^+(x,t))
     \,\le\, \inf(\|\tilde w_1(t)\|_{X^1}, \|\tilde w_1^+(t)\|_{X^1}) \cr
   \,&\le\,  N \inf (\|(\phi_0,\phi_1)\|_{Z_\epsilon^1},
     \|(\phi_0^+,\phi_1^+)\|_{Z_\epsilon^1}) \,\le\, K~,}
$$
for all $(x,t) \in \real \times [0,T)$. Therefore, by \clm(41),
$$
  w(x,t) \,\ge\, -(1 - d)a(x)^{-1} g(x)~, \quad \forall\, (x,t) \in \real
  \times [0,T)~,
$$
which allows us to apply \clm(d=zero). It follows that
$$
  \|(w(t),w_t(t))\|_{Z_\epsilon^1} \,\le\, K_7 (1 + e^{\rho T})
  \| (\phi_0, \phi_1) \|_{Z_\epsilon^1} (1 + \|\phi_0\|_{X^0})^{1/2}~,
\quad \forall\, t \in [0,T)~, \EQ(441)
$$
which contradicts \equ(436). Thus $T = \infty$. The estimates \equ(434),
\equ(435) are direct consequences of \clm(41), \clm(43) and \clm(44).
If $d > 0$, the properties \equ(tozero), \equ(decay) are obtained like in
the proof of \clm(local) in Section 3. \QED

\REMARK If $d=0$, we can still show, by arguing as in the proof of
\clm(local) in Section 3, that
$$
  \lim_{t \to +\infty} \left( \| w_x(t)\|_{L^2} + \|w_t(t)\|_{L^2} \right)
  \,=\, 0~,
$$
and that this quantity is $\OO(t^{-1/4})$ if $c=2$. However, since
\equ(dotf2) no longer holds, we cannot show that
$\|(w,w_t)\|_{Y_{\epsilon a}}$ converges to zero in this case.

\CLAIM Proposition(47) Under the assumptions of \clm(46), the solution
$(w,w_t) \in  C^0(\real_+, Z_{\epsilon}^1)$ of \equ(w) with initial
data $(\phi_0, \phi_1)$ satisfies
$$
  \lim_{t \to +\infty} t^{1/4}  \|w(t)\|_{L^\infty} \,=\, 0~. \EQ(442)
$$
If, in addition, $d>0$ and $\epsilon < 1/4$, then
$$
  \lim_{t \to +\infty} t^{1/4} \left( \| w(t)\|_{L^\infty}
  + \|(w(t),w_t(t))\|_{Y_{\epsilon a}} \right) \,=\, 0~. \EQ(443)
$$

\PROOF We first prove \equ(442). From \equ(435), it follows that
$$
  \|w(t)\|_{L^\infty} \,\le\, \sup( \|\tilde w_{-1}^+(t)\|_{L^\infty},
  \|\tilde w_{-1}^-(t)\|_{L^\infty})~, \quad \forall t\in \real_+~.\EQ(444)
$$
Therefore, we need only show that \equ(442) holds for any solution
$\tilde w \in C^0(\real_+,X^1) \cap C^1(\real_+,X^0)$ of the linear equation
with constant coefficients $L_{-1} \tilde w = 0$. Again, this can be done
using the energy estimates of Section~2. Indeed, setting $\tilde w(x,t) =
\omega(x+\nu t,t)$, where $\nu = \sqrt{c^2 -4}/(1+2\epsilon c\gamma)$, we
see that $\omega(x,t)$ satisfies
$$
   \epsilon \omega_{tt} + (1+2\epsilon c\gamma)\omega_t - 2 B \omega_{xt}
   \,=\, A \omega_{xx}~, \EQ(446)
$$
where $B > 0$ and $A = (1 + \epsilon c^2 + \epsilon c \sqrt{c^2 - 4})/
(1+ 2\epsilon c \gamma) > 0$. Remark that the coefficient of $\omega_x$
vanishes in \equ(446), like for the equation \equ(w) in the case $c=2$.
Therefore, setting
$$
  E_0(t) \,=\, \int_{\real}\left({\epsilon\over 2} \omega_t^2 +
  {1 \over 2} A|\omega_x|^2 \right)\, dx ~,
$$
and modifying accordingly the definitions of $E_1(t)$ and $E_2(t)$, we show
like in Section~2 that $\dot E_0(t) \le 0$ and that \equ(bde2), \equ(dote2)
hold. Then arguing like in Section 3, we find that $\|\omega(t)\|_{H^1} \le
C_0\|\omega(0)\|_{H^1}$, and $\lim_{t \to +\infty} t^{1/2}
\|\omega_x(t)\|_{L^2} = 0$. Since $\|\omega(t)\|_{L^\infty} \le
\|\omega(t)\|_{L^2}^{1/2} \|\omega_x(t)\|_{L^2}^{1/2}$, we see that
$\lim_{t \to +\infty} t^{1/4}\|\omega(t)\|_{L^\infty} = 0$,
which together with \equ(444) proves \equ(442).

To prove \equ(443), we recall that, if $\epsilon < 1/4$, we can define
$F_2(t)$ by \equ(f2) with $\beta = 0$. Indeed, if $\epsilon = 1/4-\delta$
for some $\delta > 0$, it is easy to verify that, under the assumption H2,
$$
  F_2(t) \,\ge\, \delta K_3(c) \|W(t)\|_{Y_{\epsilon a}}^2~,\EQ(447)
$$
for some constant $K_3(c) > 0$. Proceeding again like in the proof of
\clm(f2), we show that
$$
   \dot F_2(t) + \kappa F_2(t) \,\le\, C_1 \int_{x \ge x_d} a^2(x)
   w^2(x,t) dx \,\le\, C_2 \|w(t)\|_{L^\infty}^2~, \EQ(448)
$$
where $C_2 = (C_1/2\gamma)e^{-2\gamma x_d}$. Integrating this
differential inequality and using \equ(442) and \equ(447), we obtain
\equ(443). The proof of \clm(47), hence of \clm(decay), is complete.
\QED

\REMARK Since $L_{-1}$ is a linear operator with constant coefficients,
it is possible to obtain explicit expressions for the solutions of the
equation $L_{-1} \tilde w = 0$ in terms of the initial data, see for
example [Sm], chap.~VII.2.5. Therefore, \equ(442) could also be proved
by a direct (but cumbersome) calculation.

\SECTION The Limiting Case $\epsilon=0$

\vskip -\beforesectionskipamount
\SUBSECTION Local Stability

If we set $\epsilon=0$ in \equ(u), we obtain the well-known parabolic
KPP equation, the travelling wave solutions $g(x)$ of which are given
by \equ(front) for $c \ge 2$. To study their stability, we proceed
like in the Introduction. First, using the change of variables \equ(cv),
we arrive at \equ(v) with $\epsilon=0$. Then, we look for solutions of the
form $v(x,y,t) = g(x) + a(x)w(x,y,t)$, where $a(x) = e^{-\gamma x}$, and
we are led to study the stability of the solution $w=0$ of the parabolic
equation \equ(ww) for $\epsilon=0$ in the Sobolev space
$X^1 \equiv H^1 \cap H^1_a$. Again, linear stability holds if and only if
$1 - c\gamma + \gamma^2 \le 0$, so the biggest perturbation space is
obtained by choosing $\gamma$ as in \equ(gamma). Then, the equation \equ(ww)
for $\epsilon=0$ becomes
$$
   w_t \,=\, w_{xx} + \Delta_y w + \sqrt{c^2-4} \,w_x - 2gw -aw^2~.
   \EQ(parabo)
$$
It is known in this case that the origin is stable in $X^1$, with
polynomial decay of the perturbations to zero as $t \to +\infty$.

\REMARK In the case $c>2$, it is also known that the origin is exponentially
stable in $X^1$ if $1 - c\gamma + \gamma^2 < 0$. The best decay rate is
obtained for the value $\gamma = c/2$ [Sa], which is precisely \equ(hatgam)
for $\epsilon = 0$.

In Section~2, we have introduced various energy functionals for $\epsilon >0$,
which were used to estimate the different norms of the solution
$(w, w_t)$ of \equ(w). These functionals are all well defined for $\epsilon=0$
and allow us to control the norm of the solution $w$ of \equ(parabo).
Since all the estimates are uniform in $\epsilon$ as
$\epsilon$ goes to $0$, we can follow exactly the lines of the proof of
\clm(local), and we arrive at the (already known) local stability result:

\CLAIM Theorem(paralocal) Assume that $n \le 4$, and $c \ge 2$. Then there
exist constants $\delta_0 > 0$ and $K_0 > 0$ such that the following holds:
for all $\phi_0 \in X^1$ satisfying $\|\phi_0\|_{X^1} \le \delta_0$, there
exists a unique solution $w \in C^0(\real_+,X^1)$ of \equ(parabo) with initial
condition $w(0) = \phi_0$. Moreover, one has $\|w(t)\|_{X^1} \le K_0
\|\phi_0\|_{X^1}$ for all $t \ge 0$, and
$$
   \lim_{t \to +\infty} \left( \|\nabla w(t)\|_{X^0} +
   + \|w(t)\|_{L^2_a} \right) \,=\, 0~.
$$
In addition, if $c=2$, one has
$$
   \lim_{t \to +\infty} \sqrt{t} \left( \|\nabla w(t)\|_{X^0}
   + \|w(t)\|_{L^2_a} \right) \,=\, 0~.
$$

\REMARK Contrary to the hyperbolic case, a decay rate in time of the
solution $w(t)$ of \equ(parabo) is easily obtained for all $c \ge 2$.
Indeed, following the ideas of Nash, it is a classical task to estimate
the $L^p$-norm of solutions to parabolic equations for $p \ge 2$.
In our case, if we know an upper bound on $\|w(t)\|_{L^p}$, then we can show
that $\|w(t)\|_{L^{2p}}$ decays to zero (like an inverse power of $t$)
as $t \to \infty$, see [FS]. Thus, using the $L^2$-bound of \clm(paralocal)
and proceeding by recursion, we can show that
$$
   \|w(t)\|_{L^q} + \|w(t)\|_{L^q_a}  \,=\, \OO\left(t^{-\eta}\right)~,
   \quad t \to +\infty~,
$$
where $\eta = n(q-2)/(4q)$ and $q > 2$ is as in \equ(qdecay).

\SUBSECTION Global Stability

Like in the hyperbolic case, we obtain a global stability result when
$n = 1$. But here we apply the Maximum Principle for
parabolic equations on unbounded domains as given in [PW], Section~3.6.
Remark that it is required that $w(x,t)$ does not grow faster than
$\exp(Cx^2)$ as $x$ goes to $\pm \infty$ (uniformly in $t$), a condition
which is clearly satisfied in our case.
Like in Paragraph~4.2, we denote by $\Sigma_d(t) \in \LL(X^1,X^1)$ the
linear semigroup associated with the equation
$$
   w_t \,=\, w_{xx} + \sqrt {c^2 - 4}\,w_x - (1 +d )g w~, \quad
   d \in [-1,1]~.
$$
For $\varphi_0 \in X^1$, we set $\Sigma_d(t) \varphi_0 = w_d(t)$,
$\Sigma_d(t) \varphi_0^\pm = w_d^\pm(t)$, where $\varphi_0^\pm$ have been
given in \equ(phi+), \equ(phi-). Then, following the lines of the proofs
contained in Section~4, we obtain the global stability result below,
which has already been known, though maybe not exactly in this form.

\CLAIM Theorem(paraglobal) Assume that $n=1$, and let $c \geq 2$,
$d \in (0,1]$. Then, for any $\phi_0 \in X^1$ satisfying \equ(phi0),
namely $\phi_0(x) \ge -(1-d)a(x)^{-1}g(x)$ for all $x \in \real$,
there exists a unique solution $w\in C^0(\real_+,X^1)$ of \equ(parabo)
with initial condition $w(0) = \phi_0$. Moreover, one has
$w(x,t) \ge -(1-d)a^{-1}(x)g(x)$, and
$$
  w_{-1}^-(x,t)\, \leq\, w(x,t) \,\le\, w_1(x,t)\, \leq \,w_{-1}^+(x,t)~,
$$
for all $x \in \real$, $t \in \real_+$. In particular, if $d>0$,
$$
   \lim_{t \to + \infty} t^{1/4} \left(\|w(t)\|_{L^\infty} +
   \|w(t)\|_{H_a^1} \right) \,=\, 0~.
$$

\null
\APPENDIX(A) Maximum Principle for a Hyperbolic Operator

We consider the following hyperbolic operator $\LL$ with constant
real coefficients
$$
  \LL(u) \, = \, A u_{xx} + 2 B u_{xt} + C u_{tt} + D u_x + E u_t~,
$$
where
$$
  C \,<\, 0~, \quad B^2 - AC \,>\, 0~. \EQ(A.1)
$$
We introduce a function $h\in C^0(\real\times [0,T])$ satisfying
$$
  h(x,t) \,\ge\, \underline h, \quad \hbox{for all }(x,t)\in\real
  \times [0,T]~,\EQ(A.2)
$$
where $T$ is a positive number and $\underline h$ is a real number.
We suppose in addition, that the condition
$$
  (E^2 - 4C \underline h) (B^2 - AC) \,\ge\, (BE - CD)^2~, \EQ(A.3)
$$
holds. Finally, we set $L\, = \, \LL + h(x,t)$.
The following Maximum Principle is a simple consequence of the one given
by Protter and Weinberger (see [PW], Chapter~4, Theorem~1).

\CLAIM Theorem(MP) Assume that the conditions \equ(A.1), \equ(A.2) and
\equ(A.3) hold. If the function $(u(x,t), u_t(x,t))$ belongs to
$C^0([0,T], H_{loc}^1(\real) \times L_{loc}^2(\real))$,
with $Au_{xx} + 2Bu_{xt} + Cu_{tt}$ in $L_{loc}^2 (\real \times (0,T))$,
and satisfies the following properties,
$$
  L(u)(x,t) \ge 0~, \quad \pp(x,t) \in \real \times [0,T]~, \EQ(A.4)
$$
$$
  u(x,0) \le 0~, \quad \forall \,x \in \real~, \EQ(A.5)
$$
$$
  -C u_t(x,0) - B u_x (x,0) -{1\over 2} E u(x,0) \le 0~,\quad \pp~,
  \EQ(A.6)
$$
then $u(x,t) \le 0$ for all $(x,t) \in \real\times [0,T]$.

\PROOF Protter and Weinberger proved their Maximum Principle under the
stronger assumption $u(x,t) \in C^2(\real \times (0,T)) \cap
C^1(\real \times [0,T))$, but their proof generalizes easily to
functions $u$ satisfying the weaker regularity hypothesis
$(u(x,t), u_t(x,t)) \in C^0([0,T], H_{loc}^1(\real)\times L_{loc}^2(\real))$,
with $Au_{xx} + 2Bu_{xt} + Cu_{tt}$ in $L_{loc}^2 (\real \times (0,T))$.
Indeed their key identity (see [PW], Equation (3), page 202) still holds under
these weaker regularity assumptions and is proved by a density argument.

If $E=D=0$, the result of \clm(MP) is a direct consequence of the above remark
and of Theorem~1 of [PW]. Indeed, thanks to our assumptions \equ(A.1),
\equ(A.2), \equ(A.3), the condition of [PW] on the operator $L$, that is
$h(x,t) \ge 0$, is clearly satisfied. Since $E=0$, the conditions required
on $u(x,t)$ are exactly the hypotheses \equ (A.4) to \equ (A.6).

If $E \neq 0$ or $D \neq 0$, we reduce our problem to the previous case by
introducing the function
$$
  v(x,t) \,=\, e^{-\alpha t -\beta x} u(x,t)~,
$$
where
$$
  \alpha\,=\,{EA -BD\over 2(B^2 - AC)}~, \quad
  \beta\,=\,{CD - EB\over 2(B^2 - AC)}~.
$$
A short computation shows that
$$
  L(u)(x,t) \,=\, e^{\alpha t + \beta x} \tilde L(v)(x,t)~,
$$
where $\tilde L(v)\, =\,Av_{xx} + 2Bv_{xt} + Cv_{tt} + \tilde h v$ and
$$
  \tilde h(x,t) \ =\, -{1\over 4C} \bigl\{ (E^2 - 4Ch(x,t)) -
  {(EB-CD)^2\over B^2-AC}\bigr\}~.
$$
Now, we can apply the previous Maximum Principle, where $D=E=0$, to the
operator $\tilde L$ and to the function $v$. Indeed, due to the hypotheses
\equ (A.1), \equ(A.2), \equ (A.3), $\tilde h(x,t) \ge 0$ for all $(x,t) \in
\real\times [0,T]$. Moreover, $\tilde L(v)(x,t) \ge 0~\pp(x,t) \in \real
\times [0,T]$ and $v$ satisfies
$$
  v(x,0) \,\le\, 0~, \quad -C v_t(x,0) - B v_x (x,0) \,\le\, 0~,
$$
which are exactly the required conditions. Thus, we have proved that
$v(x,t) \le 0$, hence $u(x,t) \le 0$ for all $(x,t) \in \real\times [0,T]$.
\QED

\REMARK \clm(MP) suggests the following definition of a partial order
in $H^1_{loc}(\real) \times L^2_{loc}(\real)$. We say that $(\phi_0,\phi_1)
\le (\psi_0,\psi_1)$ if
$$ \eqalign{
   \phi_0(x) \,&\le\, \psi_0(x)~, \quad \forall\,x \in \real~, \cr
   -C \phi_1(x) - B \phi_0'(x) - {1 \over 2} E \phi_0(x) \,&\le\,
   -C \psi_1(x) - B \psi_0'(x) - {1 \over 2} E \psi_0(x) \quad \pp~,}
$$
see \equ(A.5), \equ(A.6). Then, if $(\phi_0,\phi_1) \le (\psi_0,\psi_1)$,
the solution of the linear hyperbolic equation $L(u)(x,t) = 0$ satisfying
$u(x,0) = \phi_0(x)$,
$u_t(x,0) = \phi_1(x)$ stays for all $t \in \real_+$ below the corresponding
solution with initial data $(\psi_0,\psi_1)$. An important property of this
order is that we can write any $(\phi_0,\phi_1) \in H^1_{loc} \times
L^2_{loc}$ as the sum of a ``positive'' part $(\phi_0^+,\phi_1^+) \ge 0$ and
a ``negative'' part $(\phi_0^-,\phi_1^-) \le 0$. This decomposition is unique
if we impose that $(\phi_0^+,\phi_1^+) = 0$ whenever $(\phi_0,\phi_1) \le 0$
and $(\phi_0^-,\phi_1^-) = 0$ whenever $(\phi_0,\phi_1) \ge 0$.
In the case of the operator $L_d$ defined in \equ(ld), for which
$C = -\epsilon$, $B = \epsilon c$, $E = -(1+2\epsilon c\gamma)$, the
corresponding formulae for $(\phi_0^\pm,\phi_1^\pm)$ are given in \equ(phi+),
\equ(phi-).
\def\pap{\rm }
\def\bok{\sl }

\ACKNOWLEDGEMENTS This work has been partially supported by the
Fonds National Suisse (Th.G.) and the CNRS (G.R.).

\REFERENCES

\item{[AW]} D.G. Aronson and H.F. Weinberger:
  {\pap Multidimensional Nonlinear Diffusion Arising in Population Genetics},
  Adv. in Math. {\bf 30} (1978), 33--76.

\item{[Bn]} M. Bramson:
  {\bok Convergence of Solutions of the Kolmogorov Equation to Travelling
  Waves}, Memoirs of the AMS {\bf 44}, nb. 285, Providence (1983).

\item{[Br]} P. Brenner:
  {\pap On $L_p - L_{p'}$ Estimates for the Wave Equation},
  Math. Z. {\bf 145} (1975), 251--254.

\item{[BK]} J. Bricmont and A. Kupiainen:
  {\pap Stability of Moving Fronts in the Ginzburg-Landau Equation},
  Comm. Math. Phys. {\bf 159} (1994), 287--318.

\item{[DO]} S.R. Dunbar and H.G. Othmer: {\pap On a Nonlinear Hyperbolic
  Equation Describing Transmission Lines, Cell Movement, and Branching
  Random Walks}, in {\bok Nonlinear Oscillations in Biology and Chemistry},
  H.G. Othmer (Ed.), Lect. Notes in Biomathematics {\bf 66}, Springer (1986).

\item{[EW]} J.-P. Eckmann and C.E. Wayne:
  {\pap The Nonlinear Stability of Front Solutions for Parabolic Partial
  Differential Equations}, Comm. Math. Phys. {\bf 161} (1994), 323--334.

\item{[Fa]} E.B. Fabes and D.W. Stroock:
  {\pap A New Proof of Moser's Parabolic Harnack Inequality Using the
  Old Ideas of Nash}, Arch. Rat. Mech. Anal. {\bf 96} (1986), 327--338.

\item{[Fi]} R.A. Fisher:
  {\pap The Advance of Advantageous Genes}, Ann. of Eugenics {\bf 7} (1937),
  355--369.

\item{[Ga]} Th. Gallay:
  {\pap Local Stability of Critical Fronts in Nonlinear Parabolic
  Partial Differential Equations}, Nonlinearity {\bf 7} (1994), 741--764.

\item{[Ha]} K.P. Hadeler:
  {\pap Hyperbolic Travelling Fronts}, Proc. Edinb. Math. Soc. {\bf 31}
  (1988), 89--97.

\item{[Ka]} T. Kapitula:
  {\pap On the Stability of Travelling Waves in Weighted $L^\infty$
  Spaces}, J. Diff. Eqns {\bf 112} (1994), 179--215.

\item{[Ki]} K. Kirchg\"assner:
  {\pap On the Nonlinear Dynamics of Travelling Fronts}, J. Diff. Eqns.
  {\bf 96} (1992), 256--278.

\item{[KPP]} A.N. Kolmogorov, I.G. Petrovskii and N.S. Piskunov:
  {\pap Etude de la diffusion avec croissance de la quantit\'e de mati\`ere
  et son application \`a un probl\`eme biologique}, Moscow Univ. Math. Bull.
  {\bf 1} (1937), 1--25.

\item{[KR]} K. Kirchg\"assner and G. Raugel:
  {\pap Stability of Fronts for a KPP system : The Non-Critical
  Case}, preprint (1996).

\item{[Li]} J.-L. Lions:
  {\bok Quelques m\'ethodes de r\'esolution des probl\`emes aux limites
  non lin\'eaires}, Paris, Dunod (1969).

\item{[MJ]}J.-F. Mallordy and J.-M. Roquejoffre:
  {\pap A Parabolic Equation of the KPP type in Higher Dimensions},
  SIAM J. Math. Anal. {\bf 26} (1995), 1--20.

\item{[Pa]} A. Pazy:
  {\bok Semigroups of Linear Operators and Applications to Partial
  Differential Equations},
  Appl. Math. Sci. {\bf 44}, Springer, New-York (1983).

\item{[PW]} M.H. Protter and H.F. Weinberger:
  {\bok Maximum Principles in Partial Differential Equations},
  Prentice Hall, Englewood Cliffs N.J. (1967).

\item{[RK]} G. Raugel and K. Kirchg\"assner:
  {\pap Stability of Fronts for a KPP system: The Critical
  Case}, in preparation.

\item{[Sa]} D.H. Sattinger:
  {\pap On the Stability of Waves of Nonlinear Parabolic Systems},
  Adv. Math. {\bf 22} (1976), 312--355.

\item{[Sm]} V. Smirnov:
  {\bok Cours de Math\'ematiques Sup\'erieures},
  vol. II, MIR, Moscow (1979).

\item{[St]} R. Strichartz:
  {\pap Restrictions of Fourier Transforms to Quadratic Surfaces and
  Decay of Solutions of Waves Equations}, Duke Math. J. {\bf 44} (1977),
  705--714.

\end